\PassOptionsToPackage{table}{xcolor}
\documentclass{article} 
\usepackage{iclr2027_conference,times}
\usepackage{graphicx}

\usepackage{amsmath,amsfonts,bm}

\def\eqref#1{equation~\ref{#1}}

\def\1{\bm{1}}

\DeclareMathAlphabet{\mathsfit}{\encodingdefault}{\sfdefault}{m}{sl}
\SetMathAlphabet{\mathsfit}{bold}{\encodingdefault}{\sfdefault}{bx}{n}

\usepackage{hyperref}
\usepackage{url}
\usepackage{booktabs}
\usepackage{array}
\usepackage{multirow}
\usepackage{enumitem}
\usepackage{placeins}
\usepackage{colortbl}
\usepackage{pifont}


\newcolumntype{L}[1]{>{\raggedright\arraybackslash}p{#1}}
\newcolumntype{C}[1]{>{\centering\arraybackslash}p{#1}}

\title{Physiology as Topology: Intervenable Organ Graphs for Longitudinal Patient Forecasting}

\author{Xinru Liang \\
College of AI \\
Tsinghua University \\
\texttt{liangxr25@mails.tsinghua.edu.cn} \\
\And
Yongchao Chen \thanks{Corresponding author.}\\
College of AI \\
Tsinghua University \\
\texttt{chenyongchao@mail.tsinghua.edu.cn}
}

\iclrfinalcopy

\begin{document}

\maketitle
\lhead{}

\begin{abstract}
Do language models forecast patient trajectories by reasoning through physiology, or by fitting surface statistics? We introduce PhysioCircuit, a multi-agent agent system that represents each organ as an agent and each physiological relation as an editable mechanism card with direction, lag, and evidence. Its expert-defined graph is frozen at inference, enabling controlled edge interventions. We also introduce PhysioCircuit-Bench, a mechanism-controlled benchmark spanning three organ systems, treatment events, and real ICU admissions. We evaluate intervention-direction agreement (IDAR), which requires the affected organ to change in the expected direction within the expected lag window. The expert graph substantially outperforms all topology controls on edge-IDAR, including fully connected, random, and reversed graphs. Independent clinician adjudication supports the counterfactual results. PhysioCircuit transfers to new organ systems by configuration alone, without retraining, while preserving inspectable and editable relations. Code and benchmark are available at \url{https://github.com/KatherineG11/physiocircuit}.
\end{abstract}

\section{Introduction}
\label{sec:intro}

Large language models (LLMs), built on transformer architectures and instruction tuning \citep{Vaswani2017attention, brown2020language, bommasani2021opportunities, ouyang2022training, Touvron2023llama}, now forecast patient trajectories from electronic health records \citep{achiam2023gpt, makarov2025large, li2025care, singhal2023large, Nori2023capabilities, moor2023foundation, khan2025comprehensive}. Yet predictive accuracy alone does not reveal whether a model followed the right physiology. The distinction is decisive when organs interact: a change in one organ reaches another with a specific direction and delay, and a forecaster that lacks this structure cannot be inspected, intervened on, or trusted.

The human body does not admit a precise mathematical model. Coupled organs, overlapping time scales, and pathological cross-talk defeat any equation-level simulator. We therefore build the simulation on LLM capabilities, which already integrate heterogeneous clinical knowledge and describe local state changes. The design goal is structural: an explicit physiology that constrains this flexibility and approximates real organ interactions as closely as the evidence supports. This builds on deep learning's representational foundations \citep{hinton2006fast}.

\begin{figure}[t]
  \centering
  \includegraphics[width=\linewidth]{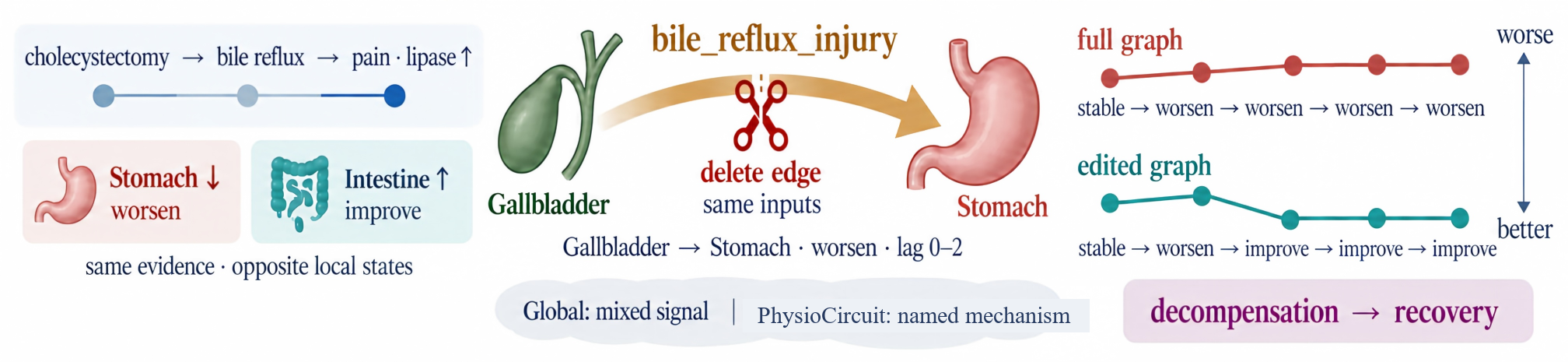}
  \caption{Same patient, one editable mechanism, different trajectory. A post-cholecystectomy patient develops bile reflux, worsening gastric injury, and improving intestinal fat absorption. Global aggregation collapses the conflicting local states into a mixed signal without exposing a named cause. PhysioCircuit routes the conflict through the editable mechanism edge \texttt{bile\_reflux\_injury}. Deleting only this edge changes the gastric trajectory from persistent worsening to improvement and shifts the macro-course from decompensation to recovery, while all other inputs remain fixed.}
  \label{fig:case_demo}
\end{figure}

Figure~\ref{fig:case_demo} shows why. After cholecystectomy, a patient develops bile reflux. Gastric injury worsens while intestinal fat absorption improves. A global predictor averages these conflicting signals into one mixed trend. PhysioCircuit keeps them separate, and the conflict runs through one editable edge, \texttt{bile\_reflux\_injury}. Deleting that single edge flips the gastric trajectory from worsening to improvement and moves the patient from decompensation to recovery, with every other input fixed. Details are in Appendix~\ref{app:intro_case}.

Existing systems leave this requirement unsolved. Data-driven organ simulators store physiology in learned parameters~\citep{chang2025organ, silveira2025multi, chen2024chexagent}; protocol systems reason at the global level~\citep{qu2026vital}; graph models infer couplings that carry no explicit clinical meaning~\citep{velivckovic2017graph}; rule systems compile biology into fixed equations~\citep{johnson2025human, wang2025multi}. Hallucination and unverifiable mechanism claims remain open risks in clinical LLM systems~\citep{ji2023survey}. They share one limit: predictive accuracy is their only score, and it cannot reveal whether a model used the correct mechanism.

We introduce \textbf{PhysioCircuit}, a multi-organ LLM framework that makes physiology the topology. Each organ is an agent with its own inspectable knowledge base. Expert-defined edges carry named mechanisms, and typed messages pass deterministic gates before receiving organs accept, partially accept, or reject them and apply bounded state updates. The graph remains frozen at inference, so deleting, reversing, or delaying one edge directly tests that mechanism.

We make three contributions.

\begin{enumerate}[leftmargin=*,label=(\arabic*),noitemsep,topsep=2pt,parsep=0pt]
    \item \textbf{Physiology is the topology.} PhysioCircuit follows expert-specified physiological structure when defining agent connections. Each mechanism edge is explicit, editable, intervenable, and verifiable.
    \item \textbf{Every organ is an agent with its own knowledge.} Each organ is an LLM agent backed by a dedicated, inspectable knowledge base. The framework uses off-the-shelf LLMs without gradient updates.
    \item \textbf{Runtime-adjustable dynamics.} Organ functions and the physiological topology can be reconfigured on demand to simulate new physiological and pathological processes, without retraining the LLM.
\end{enumerate}

Together, these results establish physiology as an inspectable and intervenable topology for clinical trajectory forecasting.

\begin{figure}[t]
  \centering
  \includegraphics[width=\linewidth]{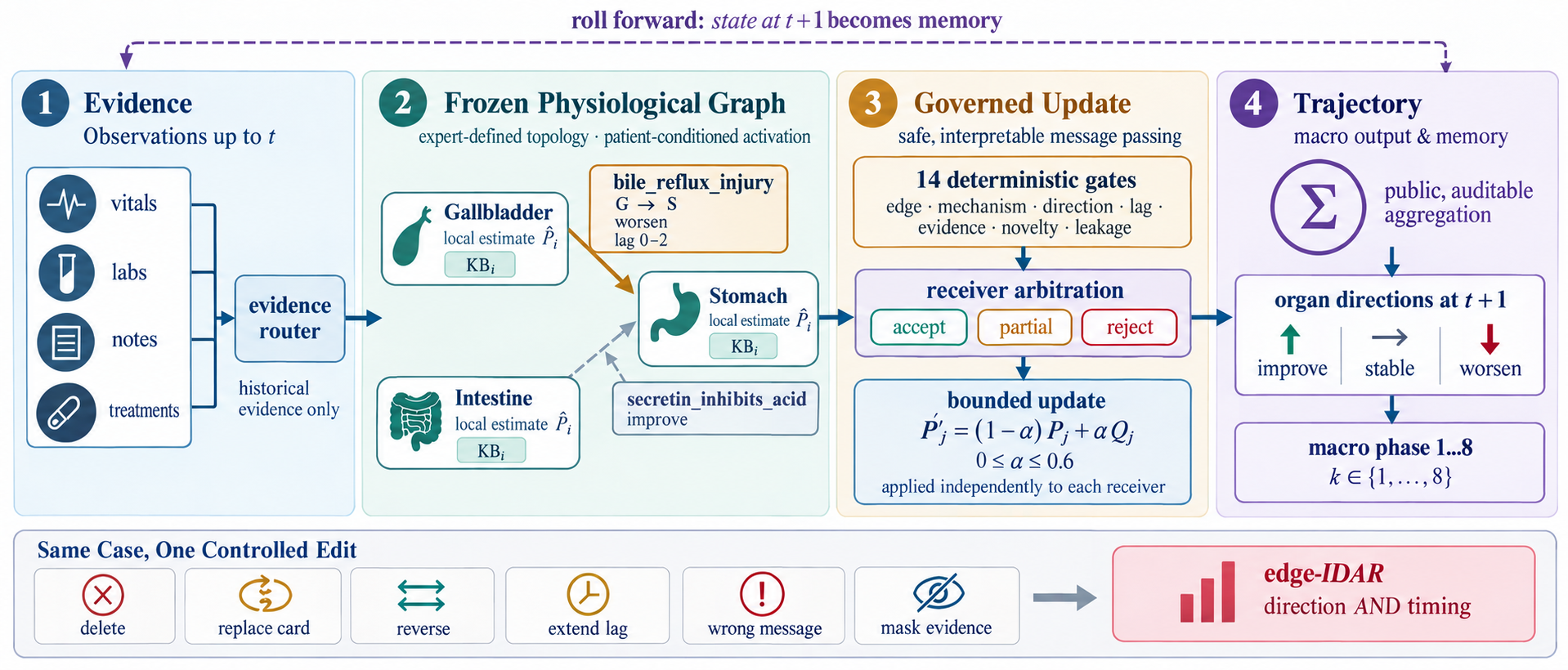}
  \caption{One inference step in PhysioCircuit. Each organ estimates locally from its own evidence; typed mechanism messages cross a frozen, expert-written organ graph under fourteen deterministic gates; surviving messages are arbitrated (accept / partial / reject) and applied through a bounded residual update; a public deterministic aggregation maps revised organ states to per-organ directions and one of eight macro phases. The graph supports intervention: every edge refers to a named mechanism card, and six operations act on one named physiological mechanism: delete edge, replace card, reverse direction, extend lag, inject wrong message, or mask source evidence.}
  \label{fig:architecture}
\end{figure}
\section{Related Work}
\label{sec:related}

Clinical simulation falls into five lines. Trajectory models forecast from EHR sequences \citep{makarov2025large, li2025care, zhou2025llm, naveed2025comprehensive}, and their internals never expose an organ mechanism one can check or change. Organ-Agents trains organ-level simulators across nine systems \citep{chang2025organ}, yet the physiology lives in weights after fine-tuning and reinforcement learning, so a new system means a new training run and no edge can be edited at inference. Vital Trace constrains longitudinal reasoning with a global protocol \citep{qu2026vital}; propagation moves through patient-level rules, and no organ edge carries direction or lag. Graph models infer couplings from data \citep{kipf2016semi, velivckovic2017graph}, and an inferred edge names no mechanism, direction, or delay. Rule-based simulators encode biology in equations that freeze at compile time.

PhysioCircuit makes physiology the editable inference structure: organs are LLM agents, links are named mechanism cards, and every edge can be deleted, reversed, or delayed at evaluation time. Table~\ref{tab:related_positioning} positions it against these lines, and Appendix~\ref{app:related} details the comparison.

\begin{table}[!ht]
    \centering
    \small
    \setlength\tabcolsep{3pt}
    \renewcommand{\arraystretch}{1.08}
    \begin{tabular}{@{}L{2.3cm} L{2.5cm} L{3.0cm} L{2.5cm} L{2.1cm}@{}}
        \toprule
        \textbf{Framework} & \textbf{Knowledge location} & \textbf{Macro course} & \textbf{Mechanism-level intervention} & \textbf{Transfer} \\
        \midrule
        Organ-Agents & model weights & data fitting + coordination & \textcolor{red}{\textbf{\ding{55}}} & \textcolor{red}{$\uparrow$} retrain \\
        Vital Trace & protocol rules & rule-constrained reasoning & \textcolor{orange}{\textbf{\ding{51}}} partial & \textcolor{orange}{$\uparrow$} rewrite \\
        Graph learning & learned weights & learned graph readout & \textcolor{red}{\textbf{\ding{55}}} & \textcolor{red}{$\uparrow$} re-learn \\
        Cell grammar & compiled rules & equation simulation & \textcolor{red}{\textbf{\ding{55}}} & \textcolor{orange}{$\uparrow$} recompile \\
        \textbf{PhysioCircuit} & \textbf{mechanism cards} & \textbf{typed propagation + arbitration} & \textcolor{green}{\textbf{\ding{51}}} & \textcolor{green}{$\downarrow$} config only \\
        \bottomrule
    \end{tabular}
    \caption{Positioning of PhysioCircuit against the closest lines of work. The table compares where knowledge is stored, how the macro course is produced, relative support for semantic intervention, and transfer cost to a new physiological system ($\downarrow$ = lower).}
    \label{tab:related_positioning}
\end{table}

The decisive test is to intervene on these explicit edges and ask whether macro outcomes move as medicine predicts, using a two-source protocol that separates truth generation from mechanism cards (Section~\ref{sec:benchmark}).

\section{Method}
\label{sec:method}

\subsection{Problem setup}

We study longitudinal forecasting over multiple organs. A case is an ordered sequence of observations $D=\{(o_1,t_1),\ldots,(o_N,t_N)\}$, where each $o_k$ carries symptoms, laboratory results, interventions, and elapsed time. From the observed prefix, the model outputs each organ direction (improve, stable, or worsen) and one of eight patient-level phases.

The forecast runs on an expert-defined mechanism graph $\mathcal{G}=(\mathcal{V},\mathcal{E},\mathcal{M})$: nodes are organs, directed edges connect organs, and $\mathcal{M}$ attaches a mechanism card to every edge. Clinicians specify the graph and freeze it before evaluation, so each connection is a named physiological relation. A new system is obtained by changing the graph and case inputs while the inference code stays fixed. Figure~\ref{fig:architecture} summarizes one inference step.

\subsection{Mechanism cards and typed messages}
\label{sec:mechanism_cards}

A mechanism card fixes a source, a target, the variables involved, the allowed direction, the lag range, supporting evidence and counterevidence, and treatment bindings. It constrains what a mechanism can do during inference; it never supplies the answer to a case. The full schema is given in Table~\ref{tab:mechanism_card} (Appendix~\ref{app:card}).

Each organ first forecasts its own state from local evidence. When an edge activates, the source emits a typed message with sender, receiver, mechanism identifier, direction, strength, lag, and evidence IDs. Deterministic gates then check edge existence, direction, lag, evidence, duplicate detection, and verdict before delivery, and they bar future observations from the current step.

\subsection{Arbitration and patient dynamics}

The receiving organ accepts, partially accepts, or rejects each surviving message and records a reason. Let $\hat{P}_j$ be its independent estimate and $\hat{Q}_j$ be the estimate conditioned on the message. The update is bounded:
\begin{equation}
\hat{P}_j'=(1-\alpha)\hat{P}_j+\alpha\hat{Q}_j, \qquad 0\leq\alpha\leq\alpha_{\max}<1.
\end{equation}
Rejection sets $\alpha=0$; stronger acceptance permits a larger revision, while the cap prevents oscillation and over-correction across steps.

A public, deterministic aggregation maps the revised organ states to one of eight macro phases. It sits outside the gold labels, so every trajectory is reproducible; the same layer emits the uncertainty used for calibration.

\subsection{Mechanism interventions and IDAR}
\label{sec:mechanism_interventions}

Because every edge refers to a mechanism card, PhysioCircuit supports controlled intervention. We use six operations: delete a mechanism, replace its card, reverse its direction, extend its lag, inject a wrong message, or mask the source evidence. Each operation is applied to the same case with all other inputs fixed. The expected response is derived from the truth team's frozen executable generator, using the same rules that produce the gold trajectories; the mechanism team that writes the cards never sees or signs these labels. The expected response specifies both the affected organ and an admissible time window.

We report three related quantities. Direction Agreement (DA) is the fraction of interventions for which the affected organ moves in the expected direction. Timing Agreement (TA) is the fraction for which the first affected-organ change falls inside the specified window. Their joint version is intervention-direction agreement (IDAR):
\begin{equation}
\mathrm{DA}=\frac{1}{N}\sum_{i=1}^{N}\mathbb{1}[d_i=\hat d_i], \qquad
\mathrm{TA}=\frac{1}{N}\sum_{i=1}^{N}\mathbb{1}[t_i\in W_i],
\end{equation}
\begin{equation}
\mathrm{IDAR}=\frac{1}{N}\sum_{i=1}^{N}\mathbb{1}[d_i=\hat d_i \ \land\ t_i\in W_i].
\end{equation}
Here $d_i$ and $t_i$ are the observed direction and first-change time, and $\hat d_i$ and $W_i$ are the expected direction and window. A non-response fails both checks. Interventions reach the model through two channels scored by the same formulas. The graph channel edits the named card or edge; we call it edge-IDAR. The edge-less channel states the same mechanism descriptor as an input-level counterfactual while the case, model, and decoding stay fixed; we call it input-IDAR.

\subsection{Inference procedure}

Each rollout follows a fixed order: local organ forecasting $\rightarrow$ typed message emission $\rightarrow$ deterministic gating $\rightarrow$ arbitration with bounded updates $\rightarrow$ public macro-phase aggregation, after which the state rolls forward. Every message, evidence item, verdict, and update is logged, making each trajectory inspectable and reproducible apart from LLM sampling. Mechanism-card fields are listed in Appendix~\ref{app:card}; gate checks are detailed in Section~\ref{sec:mechanism_cards} and the six intervention operations in Section~\ref{sec:mechanism_interventions}.

\section{PhysioCircuit-Bench: The Mechanism-Controlled Benchmark}
\label{sec:benchmark}

\subsection{Design and independence}

Standard longitudinal records describe what happened, but they do not identify which mechanism caused a transition or what would happen if that mechanism were changed. We therefore build \textbf{PhysioCircuit-Bench}, a mechanism-controlled benchmark with three complementary components. Synthetic cases provide trajectories generated from explicit physiological rules. Treatment events test whether the same rules predict responses to physiological inputs. Real ICU trajectories test whether the framework remains useful on observed clinical data.

The synthetic component uses two independently built sources. One generates patient trajectories and labels; the other writes mechanism cards and their evidence. If the model and the labels encode the same rules, the benchmark cannot distinguish mechanism reasoning from rule memorization. The two sources stay independent, the answer key for every intervention remains with the truth team, and a perturbation probe confirms that PhysioCircuit answers from the cards it was given.

\subsection{Physiological systems}

The benchmark covers three systems with separate organ sets, mechanism graphs, and 120 independent synthetic cases per system. Digestive is the source system with three organs. Pneumonia and cardiovascular-sepsis are six-organ extensions under the same protocol. The systems share no cases. The two six-organ systems share only the heart and kidney, and the digestive system shares no organ with them, so the partial overlap isolates reusable organ structure from system-specific graph content. Table~\ref{tab:systems} summarizes the systems.

Cases are authored by a truth team independent of the mechanism-card writers, independently reviewed by three physicians (Fleiss $\kappa\approx 0.72$), and split into 20 development and 100 blind-test cases per system. A third-party audit, a below-30\% knowledge-overlap bound, and an 18-rule independence probe provide evidence that the benchmark does not reduce to rule memorization. Full details of case construction, the treatment layer, counterfactual pairs, and the independence analysis are provided in Appendix~\ref{app:benchmark}.

\begin{table}[!ht]
    \centering
    \small
    \setlength\tabcolsep{3pt}
    \renewcommand{\arraystretch}{1.08}
    \begin{tabular}{@{}L{3.3cm} c c L{2.6cm} c c c@{}}
        \toprule
        \textbf{System} & \textbf{Organs} & \textbf{Cards} & \textbf{Backbone} & \textbf{Cases} & \textbf{Treat. events} & \textbf{CF pairs} \\
        \midrule
        Digestive & 3 & 12 & Qwen3-235B-A22B-Instruct-2507 & 120 & 480 & 30 \\
        Pneumonia & 6 & 14 & Qwen3-235B-A22B-Instruct-2507 & 120 & 480 & 24 \\
        Cardiovascular-sepsis & 6 & 12 & Qwen3-235B-A22B-Instruct-2507 & 120 & 480 & 24 \\
        \bottomrule
    \end{tabular}
    \caption{The three systems of the benchmark. Cards denote the number of mechanism cards, i.e.\ the graph size to specify when the system changes. Cases, treatment events, and mechanism counterfactual pairs are reported per system. The three systems share code but no cases.}
    \label{tab:systems}
\end{table}

\subsection{Evaluation protocol}

We compare PhysioCircuit with twelve component-matched baselines under matched cases, horizons, model version, and decoding settings (temperature 0.2, $\geq 3$ runs; Table~\ref{tab:baselines}; Appendix~\ref{app:benchmark}). The evaluation asks two questions. First, do methods predict local organ directions and macro phases accurately? Second, among methods exposing a semantic intervention interface, does changing a named edge move the affected organ in the expected direction and time? Graph-bearing systems are scored by edge-IDAR on named mechanism edits; edge-less systems by input-IDAR on controlled input-level counterfactuals (Appendix~\ref{app:input_protocol}).
\section{Experiments}
\label{sec:experiments}

\subsection{Experimental setup}
\label{sec:setup}

Main experiments use Qwen3-235B-A22B-Instruct-2507; cross-backbone checks use Mistral-Large-3-675B-Instruct-2512 and DeepSeek-V4-Pro-0813. All runs use temperature $0.2$, top-$p$ $1.0$, and a 1{,}024-token limit. No model is fine-tuned. Ablations use the digestive system.

\subsection{Prediction: PhysioCircuit wins where organs conflict}

Table~\ref{tab:main} shows PhysioCircuit ahead in every system: 91.3\% on digestive, 87.8\% on pneumonia, 89.0\% on cardiovascular-sepsis. It beats the global single agent by 9.2, 9.4, and 9.4 points, optimal voting by 7.8, 7.7, and 7.7, and the strongest learned-graph baseline (L) by 3.4, 4.7, and 3.6. The ordering repeats across systems, so the gain lives in the organ structure.

The gap over the strongest baseline (L) is reliable: paired case-level bootstrap 95\% CIs are [2.1, 4.7], [3.3, 6.1], and [2.3, 4.9] (20{,}000 resamples, case as unit), and McNemar exact $p<0.01$ on discordant pairs in all three systems. All results are mean $\pm$ std over three independent runs.

\begin{table}[!ht]
    \centering
    \small
    \setlength\tabcolsep{3pt}
    \renewcommand{\arraystretch}{1.06}
    \begin{tabular}{@{}lccccc@{}}
        \toprule
        \textbf{Method} & \textbf{Digest.} & \textbf{Pneum.} & \textbf{Cardio.} & \textbf{Tokens} & \textbf{Backbone} \\
        \midrule
        Global single agent & \cellcolor{blue!23}82.1$\pm$1.2 & \cellcolor{blue!11}78.4$\pm$1.1 & \cellcolor{blue!15}79.6$\pm$1.1 & 12.3/12.5/12.5K & Qwen \\
        Single agent + RAG & \cellcolor{blue!27}83.4$\pm$1.1 & \cellcolor{blue!16}79.9$\pm$1.1 & \cellcolor{blue!20}81.1$\pm$1.1 & 14.1/14.4/14.4K & Qwen \\
        Organ + optimal voting (H0) & \cellcolor{blue!27}83.5$\pm$1.1 & \cellcolor{blue!16}80.1$\pm$1.0 & \cellcolor{blue!20}81.3$\pm$1.0 & 15.6/15.9/15.9K & Qwen \\
        Compute-matched, no graph & \cellcolor{blue!29}84.2$\pm$1.0 & \cellcolor{blue!19}80.8$\pm$1.0 & \cellcolor{blue!22}82.0$\pm$1.0 & 34.8/35.4/35.4K & Qwen \\
        Implicit graph learning (L) & \cellcolor{blue!41}87.9$\pm$0.8 & \cellcolor{blue!26}83.1$\pm$0.8 & \cellcolor{blue!33}85.4$\pm$0.8 & 18.9/19.2/19.2K & Qwen \\
        Vital Trace-style & \cellcolor{blue!34}85.6$\pm$0.9 & \cellcolor{blue!22}82.0$\pm$0.9 & \cellcolor{blue!26}83.2$\pm$0.9 & 22.4/22.8/22.8K & Qwen \\
        \textbf{PhysioCircuit (PC)} & \cellcolor{blue!50}\textbf{91.3$\pm$0.7} & \cellcolor{blue!41}\textbf{87.8$\pm$0.7} & \cellcolor{blue!45}\textbf{89.0$\pm$0.7} & 41.7/42.4/42.4K & Qwen \\
        \bottomrule
    \end{tabular}
    \caption{Overall direction accuracy (\%, mean $\pm$ std over three runs). Columns are digestive, pneumonia, and cardiovascular-sepsis. All main results use Qwen3-235B-A22B-Instruct-2507; cross-backbone checks use Mistral-Large-3-675B-Instruct-2512 and DeepSeek-V4-Pro-0813. Token entries follow the same system order.}
    \label{tab:main}
\end{table}

\begin{figure}[!ht]
  \centering
  \includegraphics[width=0.85\linewidth]{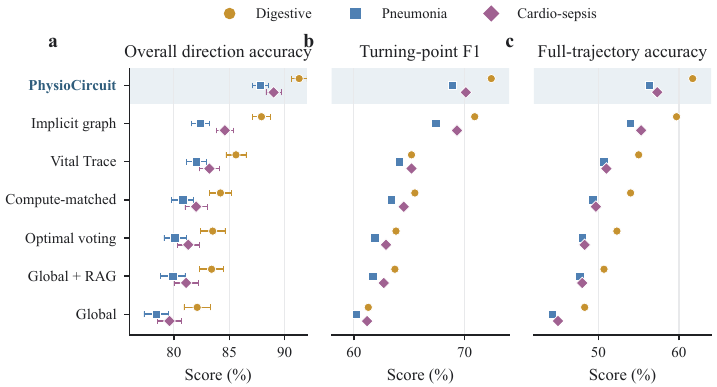}
  \caption{Predictive performance across three physiological systems. \textbf{a}, Overall direction accuracy; error bars show the reported standard deviation over three runs. \textbf{b}, Turning-point F1. \textbf{c}, Full-trajectory accuracy. Marker color and shape denote the organ system, and the shaded row highlights PhysioCircuit.}
  \label{fig:main_performance}
\end{figure}

The gain concentrates where it should. Stratified by case difficulty (Figure~\ref{fig:complexity_strata}; Appendix~\ref{app:secondary_results}), PhysioCircuit adds only 1.2, 0.8, and 0.8 points over the global agent on simple cases, where one organ dominates, and 13.6, 12.3, and 14.0 points of turning-point F1 on turning-point cases, where a macro transition needs arbitration. On conflict cases, PhysioCircuit leads implicit graph learning by 3.1 points on digestive (92.3 versus 89.2) and 3.6 on cardiovascular (87.6 versus 84.0). This is exactly the regime in which a global average folds two competing organ trends into one wrong call, whereas PhysioCircuit keeps each organ tied to its own edge. Full-trajectory accuracy, which requires every organ at every step to be correct, shows the same ordering, with PhysioCircuit leading by up to 13 points (Figure~\ref{fig:main_performance}c).

The gain is structure, not scale. The compute-matched no-graph baseline spends nearly three times the tokens (34.8K versus 12.3K for the global agent) and still trails PhysioCircuit by 7.1 points. RAG and optimal voting each recover a few points from the global agent, and neither reaches the graph.

\subsection{The topology carries the mechanism}

\begin{figure}[!ht]
  \centering
  \includegraphics[width=\linewidth]{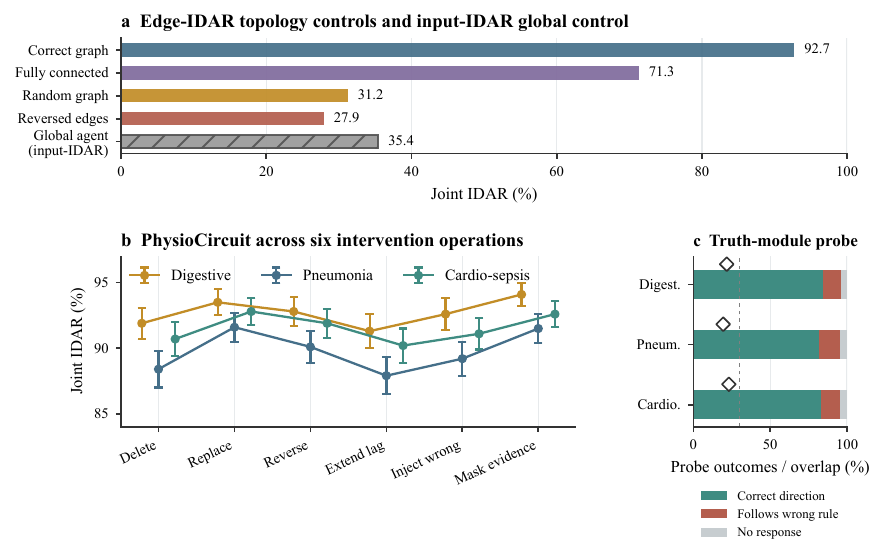}
  \caption{Semantic intervention faithfulness. \textbf{a}, Edge-IDAR for graph-bearing digestive topology controls (solid bars), with edge-less input-IDAR controls as hatched bars. A vertical divider separates the two channels. \textbf{b}, PhysioCircuit edge-IDAR across the three systems for six intervention operations. \textbf{c}, Outcome composition of the truth-module perturbation probe.}
  \label{fig:mechanism_interventions}
\end{figure}

Prediction accuracy cannot show that a model used the right organ edge. We intervene, following the causal-inference principle that a mechanism is verified by changing it and observing the effect \citep{scholkopf2021toward}. Figure~\ref{fig:mechanism_interventions}(a) puts the graph on trial. In the digestive ablation, edge-IDAR is 92.7\% for the correct expert graph, 71.3\% for a fully connected graph, 31.2\% for a random graph, and 27.9\% for a graph with reversed edges. The correct graph leads the fully connected one by 21 points, and the fully connected graph remains substantially above the random and reversed graphs. Randomizing or reversing the edges collapses the score to near-chance level. The result repeats in every system: 92.7/92.2/93.1 for the correct graph versus 29.8--32.4 for random graphs and 26.5--28.7 for reversed graphs (Table~\ref{tab:topology_all_systems}).

These expected responses carry an external clinical anchor. Three clinicians independent of the truth team adjudicated the expected direction and lag window for all 78 mechanism counterfactual pairs under blinded conditions; they reached majority consensus on 78.2\% of pairs overall and only 63.3\% of conflict pairs, and PhysioCircuit matches that independently adjudicated consensus on 91.8\% of the answerable subset (Table~\ref{tab:external_anchor}, Appendix~\ref{app:external_anchor}). The headline score is therefore not measured only against the authors' own labels.

The DA/TA decomposition pins down the failure mode (Table~\ref{tab:idar_decomposition}). The correct graph scores 94.1 for direction and 95.2 for timing. The fully connected graph drops direction to 74.2: the extra edges fire on spurious paths and misdirect the receiver. The reversed graph is the diagnostic. Its timing agreement remains at 78.4, well above its direction score (33.5) but below the correct graph (95.2) and the fully connected graph (82.1). The timing machinery partially survives; the direction is inverted. This is the cleanest evidence that the model reads the topology, not a prior about the patient.

The six intervention operations all show the expected effect (Table~\ref{tab:intervention_operations}). Delete, replace, reverse direction, extend lag, inject wrong message, and mask source evidence average 92.7/92.0/93.2 across systems. The two adversarial operations, inject wrong message and mask source evidence, score highest, 94.0/93.3/94.4 and 94.5/93.9/95.0, because they force the receiving organ to arbitrate instead of propagating blindly.

\subsection{Cards alone are not enough}

The two IDAR channels answer different questions. Edge-IDAR edits named edges on the graph and measures how sensitive the system is to topological change. On the digestive system, the correct expert graph reaches 92.7\%, a fully connected graph reaches 71.3\%, a random graph reaches 31.2\%, and a graph with reversed edges reaches 27.9\%. The closer the graph matches real physiology, the more faithful the response; this ordering holds across all three systems. This gradient shows that response quality is determined by topological quality, not by patient-level priors in the model.

Input-IDAR states the same mechanism descriptor as input text to edge-less systems and measures how much of the mechanism response can be recovered from linguistic and card knowledge alone. On input-IDAR, the global agent scores 35.4\%, 33.7\%, and 36.0\% across the three systems; optimal voting scores 38.2/36.5/39.1; no communication scores 37.0/35.8/38.3; Vital Trace-style scores 45.2/43.8/46.1; and implicit graph learning scores 58.4/56.7/59.2. The PhysioCircuit prompt-only control is an edge-less upper bound: it keeps the same LLM and card text but disables typed messages and graph propagation, reaching 62.1/59.8/63.4. Because the intervention reaches the model through different paths, the absolute values of the two channels are not directly comparable; however, the ordering within each channel is consistent across systems, and the qualitative conclusion is robust: explicit graph structure reaches a level of mechanistic faithfulness that linguistic and card knowledge alone cannot.

The independence probe confirms that the model answers from the cards it was given. When we inject 18 wrong truth rules, PhysioCircuit retains the medically correct direction on 84.6/81.5/83.2\% of probes, follows the injected rule on only 11.2/13.6/12.0\%, and reaches a joint probe IDAR of 74.5/70.4/72.8\%. A truth-retrieval control follows the injected rules 92.3\% of the time, indicating that the gap is not explained by retrieval.

\subsection{Each component earns its place}

Ablations remove one part at a time on the digestive system (Figure~\ref{fig:component_ablations}; Appendix~\ref{app:secondary_results}). Removing the mechanism cards costs 4.6 points (91.3 to 86.7); replacing typed messages with free text costs 3.4 (to 87.9); optimal voting over organ states costs 7.8. Replacing the correct graph with a random one costs 10.8 points (91.3 versus 80.5, paired CI [8.9, 12.7], $p<0.001$); reversed edges drop to 79.8; dropping organ communication drops to 81.6. A fully connected graph recovers part of the loss at 85.1 but stays 6.2 points below PhysioCircuit.

The gates do real work. The fourteen gates span six atomic checks: edge existence, direction, lag, evidence, duplicate detection, and verdict. The ablation removes four functional groups: (i) edge existence + duplicate detection, (ii) direction + lag, (iii) evidence, (iv) verdict. This cuts invalid-message interception from 87.6\% to 74.2\%, 68.9\%, 66.1\%, and 70.5\%, respectively; removing all gates raises error spread from 6.8 to 26.3.

The bounded update similarly earns its place. With $\alpha_{\max}=0.6$, five-step error growth is halved (0.42 versus 0.89) and direction flips drop from 7.1 to 3.1 per case. One-step accuracy is nearly flat across the bound, ranging from 88.8 at 0.3 to 88.4 at 1.0, so accuracy does not determine the choice; a looser bound of 1.0 accumulates error across steps.

\begin{figure}[!ht]
  \centering
  \includegraphics[width=\linewidth]{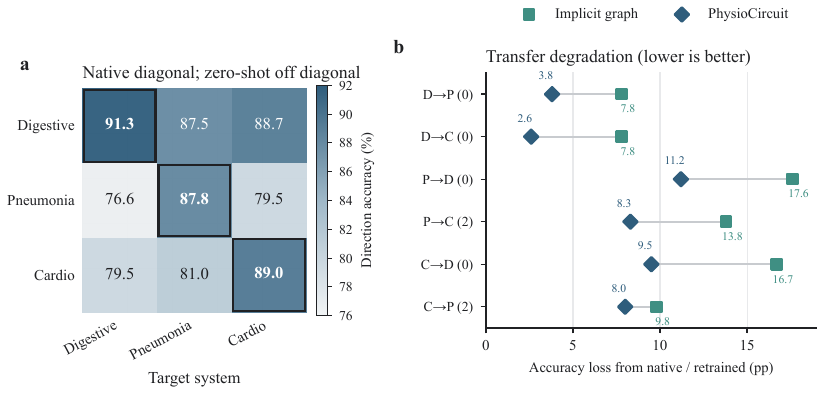}
  \caption{Cross-system transfer of explicit and learned graphs. \textbf{a}, Direction accuracy when each source graph is evaluated on each target system; outlined diagonal cells are native configuration-only runs and off-diagonal cells are zero-shot transfers. \textbf{b}, Accuracy loss relative to the native or retrained system for PhysioCircuit and implicit graph learning. Parentheses in the row labels give the number of shared organs.}
  \label{fig:cross_system_transfer}
\end{figure}

Figure~\ref{fig:cross_system_transfer} shows that the written graph transfers to a new system by configuration alone, while the implicit graph learner must relearn its couplings from data.

\subsection{Transfer, backbone, and real data}

The written graph transfers. Rewritten pneumonia and cardiovascular graphs require 127 and 154 configuration lines and reach 87.5\% and 88.7\%, only 3.8 and 2.6 points below the digestive source (Figure~\ref{fig:cross_system_transfer}). Zero-shot reuse loses 3.9 to 11.2 points and tracks organ overlap. The implicit graph learner pays 7.8 to 17.6 points to relearn its couplings, and PhysioCircuit leads it in all six transfer directions by 1.8 to 7.2 points.

The result holds across backbones. Across Qwen3, Mistral-Large-3, and DeepSeek-V4-Pro, PhysioCircuit edge-IDAR runs from 88.7 to 93.1, with per-system spread of 1.3, 3.5, and 2.8 points across the three systems, and the ordering inside every channel is unchanged (Table~\ref{tab:backbone_sweep}, Figure~\ref{fig:backbone_robustness}).

Treatment events bound to named mechanisms show the same direction and lag behavior (Figure~\ref{fig:treatment_response}, Appendix~\ref{app:secondary_results}).

Real records anchor the synthetic result. On 600 MIMIC-IV admissions \citep{johnson2023mimic}, PhysioCircuit reaches event-prediction AUROC 0.842/0.831/0.838, higher than implicit graph learning (0.834/0.821/0.827), Vital Trace-style (0.822/0.812/0.819), and the global agent (0.798/0.789/0.793) (Figure~\ref{fig:real_world_validation}(a,b) in Appendix). Mechanism-direction consistency is 73.8/71.3/74.7, and timing agreement is 70.9/69.0/71.6. These real scores sit below synthetic IDAR because co-interventions go unrecorded and timestamps are coarser, yet the ordering holds, and deterministic aggregation reaches ECE 0.0305.

\section{Conclusion}
\label{sec:conclusion}

In this work we introduced PhysioCircuit, a multi-organ LLM framework in which physiology defines the topology: each organ is an agent and every relation is an editable mechanism card carrying direction, lag, evidence, and treatment bindings. It simultaneously improves predictive accuracy and mechanism faithfulness on PhysioCircuit-Bench, three organ systems with synthetic cases, treatment counterfactuals, and a 600-admission MIMIC-IV pillar.

The correct expert graph reaches 92.7\% edge-IDAR, versus 71.3\% fully connected, 31.2\% random, and 27.9\% reversed. A prompt-only control reaches only 62\% input-IDAR, showing that card text alone cannot substitute for explicit topology. The written graph transfers to a new system by configuration alone, with no retraining. Physiology is thus an inspectable and intervenable topology for clinical trajectory forecasting.

\FloatBarrier

\subsection*{AI use statement}
LLM tools assisted language editing and programming; all numbers, experiments, clinical content, and code were reviewed by the authors. We take responsibility for the final content of this work.

\subsection*{Ethics statement}
\paragraph{Human subjects.} The 600 ICU admissions are de-identified MIMIC-IV records under its DUA, and no patient-identifiable information was retained at any stage. The three clinicians who rated trajectories under blinded conditions and independently adjudicated the 78 counterfactual expectations did so on synthetic and de-identified material. They were not members of the truth team, gave informed consent before participation, and were reimbursed at market rate, above the local hourly minimum; the adjudication task took approximately two hours per clinician.

\paragraph{Clinical use.} PhysioCircuit is a research artifact; trajectories are not cleared for clinical decision-making. The policy counterfactual is evaluated against card-defined expectations and is not treatment guidance.

\subsection*{Reproducibility statement}
We release the synthetic benchmark, mechanism cards, two-source truth module, and intervention scripts under a permissive license; the MIMIC-derived pillar cannot be redistributed but is documented for reproduction.

\bibliography{iclr2027_conference}
\bibliographystyle{iclr2027_conference}

\appendix
\setcounter{table}{0}
\renewcommand{\thetable}{A\arabic{table}}
\renewcommand{\theHtable}{A\arabic{table}}

\section{Additional Method and Benchmark Details}
\label{app:method_benchmark}

\subsection{Introductory case study}
\label{app:intro_case}

The case used in Figure~\ref{fig:case_demo} is a mechanism-conflict example from the digestive benchmark. After cholecystectomy, the patient develops bile reflux, epigastric pain, and rising lipase. The stomach shows worsening mucosal injury while intestinal fat absorption improves. A global predictor can collapse these opposing local states into a mixed signal. PhysioCircuit keeps them separate: \textsc{bile\_reflux\_injury} (gallbladder$\to$stomach, worsen, lag 0--2) explains the gastric worsening, while intestinal improvement follows a separate pathway. The macro state therefore enters \textsc{conflict\_emerges}. Table~\ref{tab:intro_case} summarizes the observed signals and mechanism-grounded reading.

\begin{table}[!ht]
    \centering
    \small
    \setlength\tabcolsep{3pt}
    \renewcommand{\arraystretch}{1.08}
    \begin{tabular}{@{}L{2.4cm} L{3.6cm} L{4.6cm}@{}}
        \toprule
        \textbf{Observed signals} & \textbf{Local organ states} & \textbf{Mechanism-grounded reading} \\
        \midrule
        Day 0: cholecystectomy. Day 3: bile reflux, epigastric pain, rising lipase & Stomach: mucosal injury worsening. Intestine: fat absorption improving & \textsc{bile\_reflux\_injury} (gallbladder$\to$stomach, worsen, lag 0--2) explains the gastric worsening; intestinal improvement is a separate pathway. Macro state enters \textsc{conflict\_emerges}. \\
        \bottomrule
    \end{tabular}
    \caption{Mechanism-conflict case used in the introductory demo. Local organ states point in opposite directions; a faithful macro-course reading composes them through typed mechanism edges with direction and lag.}
    \label{tab:intro_case}
\end{table}

In the unperturbed run, the macro course is \textsc{baseline}, \textsc{load\_accumulation}, \textsc{conflict\_emerges}, \textsc{transition}, and \textsc{decompensation}. Deleting only the \texttt{bile\_reflux\_injury} edge from gallbladder to stomach changes the gastric sequence from stable, worsen, worsen, worsen, worsen to stable, worsen, improve, improve, improve. The macro course becomes \textsc{baseline}, \textsc{load\_accumulation}, \textsc{compensation}, \textsc{recovery}, \textsc{residual\_state}. All other inputs remain fixed, isolating the effect of one editable mechanism.

\subsection{Extended related work}
\label{app:related}

\paragraph{LLMs for clinical reasoning and trajectory forecasting.}
Large language models have been used for electronic-health-record risk scoring and trajectory forecasting \citep{makarov2025large, li2025care, ren2025comprehensive, pellegriniehr2path, huang2025mimic, yang2024advancing, singhal2025toward, chen2024chexagent, silveira2025multi}. DT-GPT forecasts clinical trajectories and CARE-AD coordinates specialized agents over longitudinal notes \citep{makarov2025large, li2025care}. These systems evaluate sequence-level prediction; PhysioCircuit additionally exposes organ mechanisms for inspection.

\paragraph{Multi-agent LLM systems in medicine.}
Multi-agent collaboration supports medical debate, diagnosis, and simulation \citep{zhou2025llm, williams2023epidemic, Li2023camel, ferrag2026llm, wang2023voyager}. Retrieval, chain-of-thought, self-consistency, self-refine, reflection, and tool use provide general coordination primitives \citep{lewis2020retrieval, wei2022chain, wang2022self, madaan2023self, nye2021show, kojima2022large, Zelikman2022star, Yao2022react, Shinn2023reflexion, Schick2023toolformer, Park2023generative, hong2024metagpt, Wu2023autogen, xi2025rise}. PhysioCircuit uses these primitives with typed messages whose direction, lag, and evidence are defined by physiological mechanisms.

\paragraph{LLM-based physiological simulation.}
Text-based simulators treat LLMs as biomedical simulators without manual tuning or physics-based equations \citep{schaefer2023large, liu2026survey}. Organ-Agents trains organ-level simulators across nine systems, but the physiological knowledge lives in weights after fine-tuning and reinforcement learning \citep{chang2025organ}. Full-Body AI Agent integrates multiple biological levels but does not provide mechanism-level intervention or cross-system transfer evidence \citep{wang2025multi}.

\paragraph{Protocol-constrained reasoning.}
Vital Trace constrains longitudinal reasoning with a curated global protocol and patient-state memory, coordinated by Router, Reasoner, Auditor, and Steward agents \citep{qu2026vital}. Its state transitions are patient-level protocol rules; the unit of propagation is not a typed organ interaction with explicit direction and lag.

\paragraph{Mechanism knowledge and graph learning.}
Cell behavior grammars compile natural-language rules into mathematical models, but the compiled rules cannot be arbitrated during inference \citep{johnson2025human}. Graph convolutional and attention networks learn adjacency structures end to end from data \citep{kipf2016semi, velivckovic2017graph}; recent exploratory work also probes LLM guidance for discovering such propagation graphs \citep{hellm}. Learned edges carry coupling but no named mechanism semantics, so a user cannot delete, reverse, or delay one physiological process.

\subsection{Mechanism card fields}
\label{app:card}

\begin{table}[t]
    \centering
    \small
    \setlength\tabcolsep{3pt}
    \renewcommand{\arraystretch}{1.08}
    \begin{tabular}{@{}L{3.4cm} L{7.4cm}@{}}
        \toprule
        \textbf{Field} & \textbf{Semantics} \\
        \midrule
        \texttt{mechanism\_id} & Unique identifier of the mechanism \\
        \texttt{source}, \texttt{target} & Endpoints of the directed edge \\
        \texttt{source\_variables}, \texttt{target\_variables} & Variables involved on each side \\
        \texttt{allowed\_directions} & Whether the mechanism worsens, improves, or both \\
        \texttt{lag\_steps} & Allowed delays between activation and effect \\
        \texttt{activation\_}\\[-1pt]\texttt{evidence} & Conditions that activate the mechanism \\
        \texttt{counterevidence} & Conditions that suppress the mechanism \\
        \texttt{treatment\_bindings} & Treatments that activate or suppress the mechanism \\
        \texttt{mechanism\_}\\[-1pt]\texttt{description} & Natural-language description of the physiology \\
        \bottomrule
    \end{tabular}
    \caption{Fields of a mechanism card. A card constrains how a mechanism can act during inference; the case itself is always answered by the rollout.}
    \label{tab:mechanism_card}
\end{table}

\subsection{Implementation details}
\label{app:implementation}

\paragraph{Macro phases.} The deterministic aggregation emits one of eight phases: \textsc{baseline}, \textsc{load\_accumulation}, \textsc{conflict\_emerges}, \textsc{transition}, \textsc{decompensation}, \textsc{compensation}, \textsc{recovery}, and \textsc{residual\_state}. Phase assignment follows a fixed priority rule over per-organ directions and confidences: worsening organs with confidence above 0.6 advance the phase toward \textsc{decompensation}; stable or improving organs advance it toward \textsc{recovery}; conflicting directions across organs produce \textsc{conflict\_emerges} or \textsc{transition}. The rule is pure and stateless apart from the previous phase, so the same organ-state vector and previous phase always yield the same phase.

\paragraph{Arbitration and $\alpha$.} Each receiving organ arbitrates incoming messages into accept, partial accept, or reject. Rejection sets $\alpha=0$ (no state change); partial accept sets $0<\alpha<\alpha_{\max}$; accept sets $\alpha=\alpha_{\max}=0.6$. The bounded residual update is $\Delta s = \alpha \cdot \hat{\Delta}$, where $\hat{\Delta}$ is the message-proposed direction and magnitude.

\paragraph{Fourteen gates.} The fourteen deterministic gates span six atomic checks: edge existence (2 gates: source--target edge present, edge active at current step), direction (3 gates: direction in card-admissible set, direction not contradicted by evidence, direction monotonic with lag), lag (2 gates: current step within card lag window, lag not exceeded), evidence (3 gates: at least one evidence ID cited, evidence IDs exist in case record, no counterevidence active), duplicate detection (2 gates: same mechanism not already delivered this step, message hash not seen in current rollout), and verdict (2 gates: verdict field well-formed, verdict consistent with direction).

\subsection{Benchmark construction details}
\label{app:benchmark}

\begin{table}[t]
    \centering
    \small
    \setlength\tabcolsep{3pt}
    \renewcommand{\arraystretch}{1.06}
    \begin{tabular}{@{}C{0.7cm} L{4.2cm} L{7.6cm}@{}}
        \toprule
        \textbf{ID} & \textbf{Method} & \textbf{Rules out} \\
        \midrule
        A & Global single agent (no RAG) & organ decomposition has no value \\
        B & Single agent + mechanism RAG & gains come only from external knowledge \\
        C & Independent organs, no communication & gains are only from splitting organs \\
        D & Multi-agent, communication without gates & program gates are unnecessary \\
        E & Organ prediction + optimal voting (H0) & macro is just a combination of local states \\
        F & Compute-matched multi-agent without graph & gains are only from more calls \\
        G & Free-text multi-agent (no mechanism codes) & typed mechanism messages are unnecessary \\
        H & Random graph & correct topology is unnecessary \\
        I & Reversed / fully connected / random graph & direction and sparsity do not matter \\
        J & No gates / unbounded residual & gating and bounded update are unnecessary \\
        K & Vital Trace-style protocol system & patient-level protocol suffices \\
        L & Implicit graph learning (learned graph, GAT) & explicit mechanism cards are replaceable \\
        \bottomrule
    \end{tabular}
    \caption{The twelve baselines (A to L) and the assumption each one rules out. The three decisive contrasts are communication (independent organs vs.\ PhysioCircuit), composition (voting vs.\ PhysioCircuit), and topology (random graph vs.\ PhysioCircuit).}
    \label{tab:baselines}
\end{table}

\subsubsection{Two-source independence protocol}

Two teams construct each system from disjoint sources. The truth team writes state-transition rules from clinical guidelines and textbooks, freezes them into an executable generator, produces the gold trajectory labels, and also authors the expected direction and lag window used to score every intervention. The mechanism team writes cards from pathophysiology literature and organ-interaction reviews without seeing the truth artifact or the scoring labels. A third-party coordinator audits the split, resolves disagreements, and records the audit trail. Knowledge overlap is estimated by two independent raters and remains below the 30\% protocol bound in all systems (21.7\%, 19.4\%, and 23.1\%). Because the expected direction and lag window are still authored inside the author team, three independent clinicians separately adjudicated the same expectations under blinded conditions on the 78 mechanism counterfactual pairs; on the pairs reaching majority consensus, their expected direction agrees with the truth-team label on 95.1\% of cases (\S\ref{app:external_anchor}).

The independence probe injects $K$ medically wrong rules into the truth generator, freezes the modified generator, and regenerates a probe set. For example, the probe may assert that bile reflux improves the gastric mucosal barrier. If intervention responses follow the injected rule, the model is copying the truth generator; if they preserve the medically correct direction, the cards provide independent content. We use $K=18$ in the reported probe and require at least 80\% medically correct responses. A truth-retrieval baseline follows the injected rules on 92.3\% of cases, confirming that the probe detects truth-module mimicry.

\subsubsection{Input-level counterfactual protocol for edge-less systems}
\label{app:input_protocol}

For systems that do not expose a named organ-to-organ edge, we administer the same intervention descriptor as a controlled input-level counterfactual. The protocol holds the patient case, model, temperature, output limit, and retrieval scope fixed. The intervention descriptor is a natural-language sentence that states the mechanism, its expected direction, and its admissible lag window (e.g., ``Assume that bile no longer refluxes from the gallbladder to the stomach; the expected gastric effect is improvement within 0--2 steps.''). The model then produces a future trajectory under this descriptor, and the same DA / TA / joint IDAR scoring is applied.

Three operational definitions fix comparability. First, \emph{non-response} is defined as the affected organ's predicted trajectory being identical to its baseline run (no directional change in any step). Second, \emph{first-change time} is read from the model's step-by-step rollout; models that emit the full future sequence in a single generation are scored by reading the predicted first step that differs from baseline. Third, the same $N=180/210/190$ intervention instances are used for both edge-IDAR and input-IDAR in each system, so denominators are matched.

A PhysioCircuit prompt-only control runs the same LLM and uses the same card knowledge but disables typed messages, graph propagation, and arbitration; the intervention descriptor is injected directly into the single-prompt context. This control separates the contribution of the graph channel from those of model knowledge and card text.

\subsubsection{Synthetic cases}

Each system contributes 120 cases (20 development and 100 blind test) with five to eight future time points and three robustness variants: missing observations, noisy values, and reordered descriptions. Three physicians review each truth independently; disagreements are resolved by majority vote. Agreement by Fleiss $\kappa$ is $0.74$, $0.71$, and $0.73$ for digestive, pneumonia, and cardiovascular cases. Fingerprints over mechanism, severity, temporal pattern, evidence, intervals, and symptoms remove near-duplicates; fingerprints are computed within each system so that cases from different systems cannot collide.

\begin{table}[t]
    \centering
    \small
    \setlength\tabcolsep{3pt}
    \renewcommand{\arraystretch}{1.08}
    \begin{tabular}{@{}L{3.6cm} C{1.6cm} L{5.6cm}@{}}
        \toprule
        \textbf{Stratum} & \textbf{Share} & \textbf{Design purpose} \\
        \midrule
        Simple or near-stable & 30\% & single-organ competence \\
        Cross-organ conflict & 40\% & mechanism arbitration \\
        Turning point and decompensation & 20\% & macro transitions \\
        Intermittent and non-monotonic & 10\% & long-range structure \\
        \bottomrule
    \end{tabular}
    \caption{Stratification of the blind test set, identical in all three systems. Conflict cases carry the weight of the evaluation because they are where mechanism arbitration should matter most.}
    \label{tab:strata}
\end{table}

The blind-test strata are identical across systems (Table~\ref{tab:strata}). Gold truth, design cards, expert notes, and future time points are excluded from prompts; retrieval uses stable mechanisms and current patient memory only. Predicted nodes are marked \texttt{predicted=true}.

\subsubsection{Treatment events and treatment counterfactuals}
\label{sec:treatbench}

The treatment layer complements edge interventions with physiological inputs. Each system contributes 480 pharmacological, surgical, or supportive events, four per case. Every event binds to a card through \texttt{treatment\_bindings}, specifies the expected mechanism direction, and defines a lag window. We score both direction agreement and whether the response appears within that window. The layer also contains 360 treatment counterfactual pairs, 120 per system; each pair holds the case fixed, applies one treatment in the treated branch, withholds it in the control branch, and records a field identifying the single changed factor so that silent multi-variable edits cannot pass.

\subsubsection{Counterfactual pairs}

Each mechanism counterfactual changes one key factor and stays within its system. The digestive set has 30 pairs across five families: high-fat diet with partial biliary obstruction; bile reflux with improving fat absorption; gastric hyperacidity with later loss of intestinal compensation; transient obstruction followed by resolution; and post-cholecystectomy bile insufficiency with gastrointestinal feedback. Pneumonia has 24 pairs across airway-clearance failure, alveolar flooding with vascular loading, systemic inflammatory drive, and renal retention with uraemic airway injury. Cardiovascular-sepsis has 24 pairs across vasoplegia with myocardial depression, renal and hepatic coupling, coagulopathy with glycocalyx shedding, and recovery with restitution. Every pair records the changed factor, and no pair crosses a system boundary.

\subsubsection{External clinical anchoring of counterfactual expectations}
\label{app:external_anchor}

The expected direction and lag window used to score IDAR are authored by the truth team. To check that these expectations are not artifacts of the authors' own frozen rules, the three external clinicians from \S\ref{sec:expert} independently adjudicated all 78 mechanism counterfactual pairs. These clinicians were not members of the truth team and wrote none of the generator rules. For each pair, they saw only the de-identified case and the single changed factor; the truth-team label, the PhysioCircuit output, and the mechanism card text were hidden. They judged two quantities: the expected direction of the affected organ (worsen, improve, or no change) and the expected lag window in steps.

Table~\ref{tab:external_anchor} reports the result. Direction agreement across the three raters, measured by Fleiss $\kappa$, is 0.67 overall, 0.79 on non-conflict pairs, and 0.51 on conflict pairs. Lag-window agreement within one step is 73.1\% overall. A majority of the three raters agree on both direction and window for 61 of 78 pairs (78.2\%); on conflict pairs, where two organ pathways pull in opposite directions, majority consensus falls to 19 of 30 (63.3\%). On the 61 consensus pairs, the clinicians' majority expected direction agrees with the truth-team label on 58 of 61 (95.1\%), confirming that the scoring expectations are clinically reasonable rather than artifacts of the generator. Of the 17 pairs without majority consensus, nine disagree only on the lag window while agreeing on direction, five split on direction, and three involve competing multi-pathway responses for which the direction itself is unclear. These are intrinsically ambiguous cases, not scoring disagreements.

Restricting PhysioCircuit's edge-IDAR to the 61 consensus pairs gives 91.8\%, close to the 92.7\% headline. Where independent clinicians can agree on the expected response, PhysioCircuit recovers it; where clinicians split, the expected label is itself under-specified.

\begin{table}[t]
    \centering
    \small
    \setlength\tabcolsep{4pt}
    \renewcommand{\arraystretch}{1.08}
    \begin{tabular}{@{}lccccc@{}}
        \toprule
        \textbf{Stratum} & \textbf{N} & \textbf{Direction} & \textbf{Window} & \textbf{Majority} & \textbf{PhysioCircuit} \\
        & & \textbf{Fleiss $\kappa$} & \textbf{($\pm$1 step)} & \textbf{consensus} & \textbf{on consensus} \\
        \midrule
        Non-conflict & 48 & 0.79 & 85.4\% & 87.5\% & 93.0\% \\
        Conflict & 30 & 0.51 & 53.3\% & 63.3\% & 89.5\% \\
        \midrule
        All pairs & 78 & 0.67 & 73.1\% & 78.2\% & 91.8\% \\
        \bottomrule
    \end{tabular}
    \caption{External clinical anchoring of counterfactual expectations. Three clinicians independent of the truth team judged expected direction and lag window under blinded conditions for all 78 mechanism counterfactual pairs. Direction agreement is Fleiss $\kappa$ over three raters; window agreement allows a one-step tolerance; majority consensus requires a majority of the three raters to agree on both direction and window. On consensus pairs the clinicians' majority direction agrees with the truth-team label at 95.1\%. The last column is PhysioCircuit edge-IDAR restricted to the consensus subset.}
    \label{tab:external_anchor}
\end{table}

\subsubsection{Real longitudinal pillar}
\label{sec:realpillar}

Because real records lack mechanism-level truth, the real pillar evaluates two practice-checkable tasks: event prediction, asking whether an organ will require support, deteriorate, or recover in the coming window; and mechanism-direction consistency after a documented treatment change. It uses 600 MIMIC-IV ICU admissions \citep{johnson2023mimic}, 200 per system, selected for complete records and documented treatment changes. Each admission contributes ten treatment-change observations, yielding 2,000 direction observations per system. Scores are stratified by treatment family and include a lag-window check. This pillar anchors the synthetic intervention results in observed clinical data.

\section{Additional Experiments and Analyses}
\label{app:experiments}

\subsection{Secondary experiments}
\label{app:secondary}

\subsubsection{Treatment response and response timing}
\label{sec:treatment}

Table~\ref{tab:treatment} reports the treatment layer. Across 1,440 synthetic events, mechanism direction accuracy is 98.1\%, 97.5\%, and 98.5\% for the three systems. On 360 treatment counterfactual pairs, response direction exceeds 90\% and lag-window agreement is 88.3\%, 86.7\%, and 87.5\%. In the 6,000 real treatment-change observations, direction consistency is 73.8\%, 71.3\%, and 74.7\%; timing is 70.9\%, 69.0\%, and 71.6\%. Pharmacological changes are most consistent in all systems, surgical changes are weakest on direction, and supportive care is weakest on timing. Supportive interventions often act on several mechanisms at once, so a single-card lag window only partly describes the observed response. The synthetic layer therefore provides a controlled temporal test, while the real layer exposes the additional variability present in practice. The ablations below isolate which components support these responses.

\begin{table}[t]
    \centering
    \small
    \setlength\tabcolsep{2pt}
    \renewcommand{\arraystretch}{1.04}
    \begin{tabular}{@{}lccccc@{}}
        \toprule
        \multicolumn{6}{l}{\textit{Panel A: synthetic treatment layer}} \\
        \textbf{System} & \textbf{Events} & \textbf{Mechanism direction} & \textbf{Pairs} & \textbf{Response direction} & \textbf{Lag in window} \\
        \midrule
        Digestive & 480 & \cellcolor{orange!49}98.1\% & 120 & \cellcolor{orange!35}91.7\% & \cellcolor{orange!28}88.3\% \\
        Pneumonia & 480 & \cellcolor{orange!48}97.5\% & 120 & \cellcolor{orange!33}90.8\% & \cellcolor{orange!25}86.7\% \\
        Cardiovascular-sepsis & 480 & \cellcolor{orange!49}98.5\% & 120 & \cellcolor{orange!35}91.7\% & \cellcolor{orange!27}87.5\% \\
        \midrule
        \multicolumn{6}{l}{\textit{Panel B: real treatment-change observations (MIMIC-IV)}} \\
        \textbf{System} & \textbf{Family} & \textbf{Observations} & \multicolumn{2}{c}{\textbf{Direction consistency}} & \textbf{Lag in window} \\
        \midrule
        \multirow{3}{*}{Digestive} & pharmacological & 900 & \multicolumn{2}{c}{\cellcolor{orange!33}76.0\%} & \cellcolor{orange!25}72.2\% \\
         & supportive & 500 & \multicolumn{2}{c}{\cellcolor{orange!26}72.4\%} & \cellcolor{orange!19}67.6\% \\
         & surgical & 600 & \multicolumn{2}{c}{\cellcolor{orange!25}71.7\%} & \cellcolor{orange!25}71.7\% \\
        \midrule
        \multirow{3}{*}{Pneumonia} & pharmacological & 900 & \multicolumn{2}{c}{\cellcolor{orange!29}74.0\%} & \cellcolor{orange!22}70.0\% \\
         & supportive & 500 & \multicolumn{2}{c}{\cellcolor{orange!23}70.4\%} & \cellcolor{orange!19}67.2\% \\
         & surgical & 600 & \multicolumn{2}{c}{\cellcolor{orange!20}68.0\%} & \cellcolor{orange!21}69.0\% \\
        \midrule
        \multirow{3}{*}{Cardiovascular} & pharmacological & 900 & \multicolumn{2}{c}{\cellcolor{orange!35}77.0\%} & \cellcolor{orange!28}73.7\% \\
         & supportive & 500 & \multicolumn{2}{c}{\cellcolor{orange!27}73.2\%} & \cellcolor{orange!19}67.0\% \\
         & surgical & 600 & \multicolumn{2}{c}{\cellcolor{orange!25}72.5\%} & \cellcolor{orange!25}72.3\% \\
        \bottomrule
    \end{tabular}
    \caption{Treatment response and response timing. Panel A scores synthetic treatment events and treatment counterfactual pairs against the mechanism cards. Panel B stratifies the real treatment-change observations by treatment family and adds the timing check. Drug changes are the most consistent family, surgical changes the least consistent on direction, and supportive changes the weakest on timing in all three systems.}
    \label{tab:treatment}
\end{table}

\subsubsection{Clinical expert trajectory evaluation}
\label{sec:expert}

Expert readers assess whether a predicted course is clinically reasonable. Three clinicians rated 80 digestive trajectories from each method under blinded conditions on a five-point Likert scale (Table~\ref{tab:expert}). PhysioCircuit scores 4.31 (standard deviation 0.65), compared with 4.00 (0.72) for implicit graph learning and 3.54 (0.83) for the global agent. Exact agreement is 88.6\%, 82.3\%, and 74.3\%; Fleiss $\kappa$ for binary acceptability is 0.77, 0.76, and 0.74. The difference between PhysioCircuit and implicit graph learning is modest in accuracy but more pronounced in clinical ratings and rater agreement. Each PhysioCircuit trajectory also names the mechanisms that produced it, so a reviewer can trace a rating to a card.

\begin{table}[t]
    \centering
    \small
    \setlength\tabcolsep{3pt}
    \renewcommand{\arraystretch}{1.06}
    \begin{tabular}{@{}lcccc@{}}
        \toprule
        \textbf{Method} & \textbf{Likert mean} & \textbf{Likert std} & \textbf{Exact rater agreement} & \textbf{Fleiss $\kappa$} \\
        \midrule
        \textbf{PhysioCircuit} & \cellcolor{purple!43}\textbf{4.31} & 0.65 & \cellcolor{purple!44}88.6\% & \cellcolor{purple!38}0.77 \\
        Implicit graph learning (L) & \cellcolor{purple!40}4.00 & 0.72 & \cellcolor{purple!41}82.3\% & \cellcolor{purple!38}0.76 \\
        Global single agent & \cellcolor{purple!35}3.54 & 0.83 & \cellcolor{purple!37}74.3\% & \cellcolor{purple!37}0.74 \\
        \bottomrule
    \end{tabular}
    \caption{Clinical expert evaluation. Three clinicians rate 80 trajectories per method under blinded conditions on a five-point scale (240 ratings per method). Trajectories are drawn from the digestive system so that raters compare methods under a fixed physiological setting. Exact rater agreement is the fraction of rater pairs that give the same score to the same trajectory. Fleiss $\kappa$ is computed on the binary judgement of whether a trajectory is clinically acceptable (rating of 4 or above).}
    \label{tab:expert}
\end{table}

\subsubsection{Policy-level counterfactual}

Single-edge interventions test graph organization; a clinical decision also asks what happens when treatment is given now or later. We built 60 policy pairs, 20 per system, by applying the indicated treatment at the earliest eligible step or two to four steps later. The expected macro response is derived from the same frozen generator rules that score all other interventions, not from the cards the model reads. The early branch matches the expected response in 91.7\% of pairs and resolves in 2.40 steps, versus 48.3\% and 4.70 steps for delayed treatment (Table~\ref{tab:policy}). Matched real groups have comparable APACHE II scores \citep{apache1985apache} (18.17 vs.\ 18.90; 95\% bootstrap interval for the difference $[-2.30,0.77]$), reducing a baseline-severity explanation. This experiment tests a decision-shaped counterfactual through both response direction and response timing.

\begin{table}[t]
    \centering
    \small
    \setlength\tabcolsep{3pt}
    \renewcommand{\arraystretch}{1.06}
    \begin{tabular}{@{}lcccc@{}}
        \toprule
        \textbf{Strategy} & \textbf{Pairs} & \textbf{Expected macro response} & \textbf{Steps to resolution} & \textbf{APACHE II} \\
        \midrule
        Early intervention & 60 & 91.7\% & 2.40 & 18.17 \\
        Delayed intervention & 60 & 48.3\% & 4.70 & 18.90 \\
        \bottomrule
    \end{tabular}
    \caption{Policy-level counterfactual. Each pair holds the case fixed and moves the treatment in time. Expected macro response is the fraction of pairs whose macro outcome matches the clinically expected change. The APACHE II difference between the matched real groups is $-0.73$ with a 95\% bootstrap interval of $[-2.30, 0.77]$.}
    \label{tab:policy}
\end{table}

\subsubsection{A clinical what-if walkthrough}
\label{sec:whatif}

Case \texttt{PC-D-014} (the post-cholecystectomy example of Appendix~\ref{app:intro_case}) illustrates the audit path. In the unperturbed run, the macro course is \textsc{baseline}, \textsc{load\_accumulation}, \textsc{conflict\_emerges}, \textsc{transition}, and \textsc{decompensation}. Deleting only the \texttt{bile\_reflux\_injury} edge changes the gastric sequence from stable, worsen, worsen, worsen, worsen to stable, worsen, improve, improve, improve. The macro course becomes \textsc{baseline}, \textsc{load\_accumulation}, \textsc{compensation}, \textsc{recovery}, \textsc{residual\_state}. All other inputs remain fixed, so the example isolates the effect of one editable mechanism.

\subsection{Secondary results and ablations}
\label{app:secondary_results}

The main text reports headline numbers; all secondary results, the backbone sweep, the per-intervention-type breakdown, component ablations, cross-system transfer, the real longitudinal pillar, downstream retention, calibration, and cost are detailed below.

\begin{figure}[!ht]
  \centering
  \includegraphics[width=\linewidth]{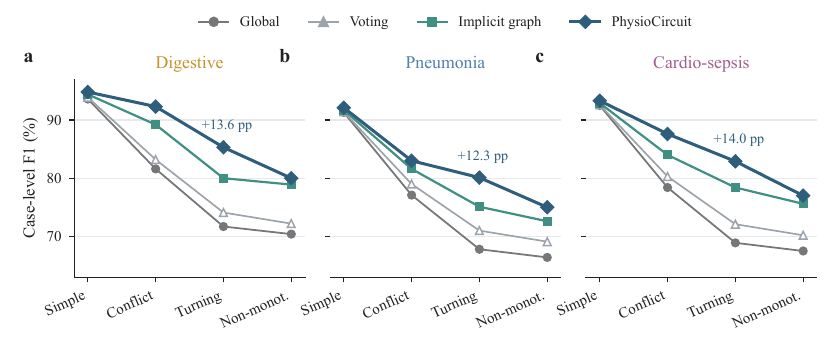}
  \caption{Performance by mechanism-complexity stratum. Direction accuracy is shown for global prediction, optimal voting, implicit graph learning, and PhysioCircuit in the digestive (\textbf{a}), pneumonia (\textbf{b}), and cardiovascular-sepsis (\textbf{c}) systems. The annotated difference is the PhysioCircuit gain over the global single agent on turning-point cases. The gain is smallest on simple cases and largest on turning-point cases.}
  \label{fig:complexity_strata}
\end{figure}

\begin{figure}[t]
  \centering
  \includegraphics[width=\linewidth]{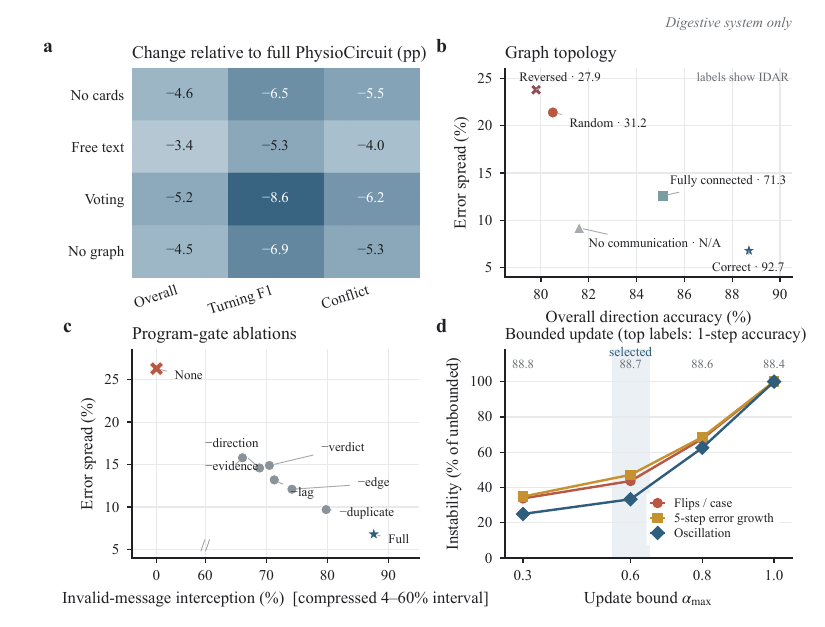}
  \caption{Component ablations on the digestive system. \textbf{a}, Change in overall accuracy, turning-point F1, and conflict accuracy relative to the full system. \textbf{b}, Overall accuracy versus error spread under graph-topology controls; edge-IDAR is reported for graph-bearing controls, while input-IDAR for the global agent is reported separately because it uses a controlled input-level counterfactual. \textbf{c}, Invalid-message interception versus error spread after removing gate categories. \textbf{d}, Flip rate, five-step error growth, and oscillation normalized to the unbounded update; numbers above the curves report one-step accuracy.}
  \label{fig:component_ablations}
\end{figure}

\begin{figure}[t]
  \centering
  \includegraphics[width=\linewidth]{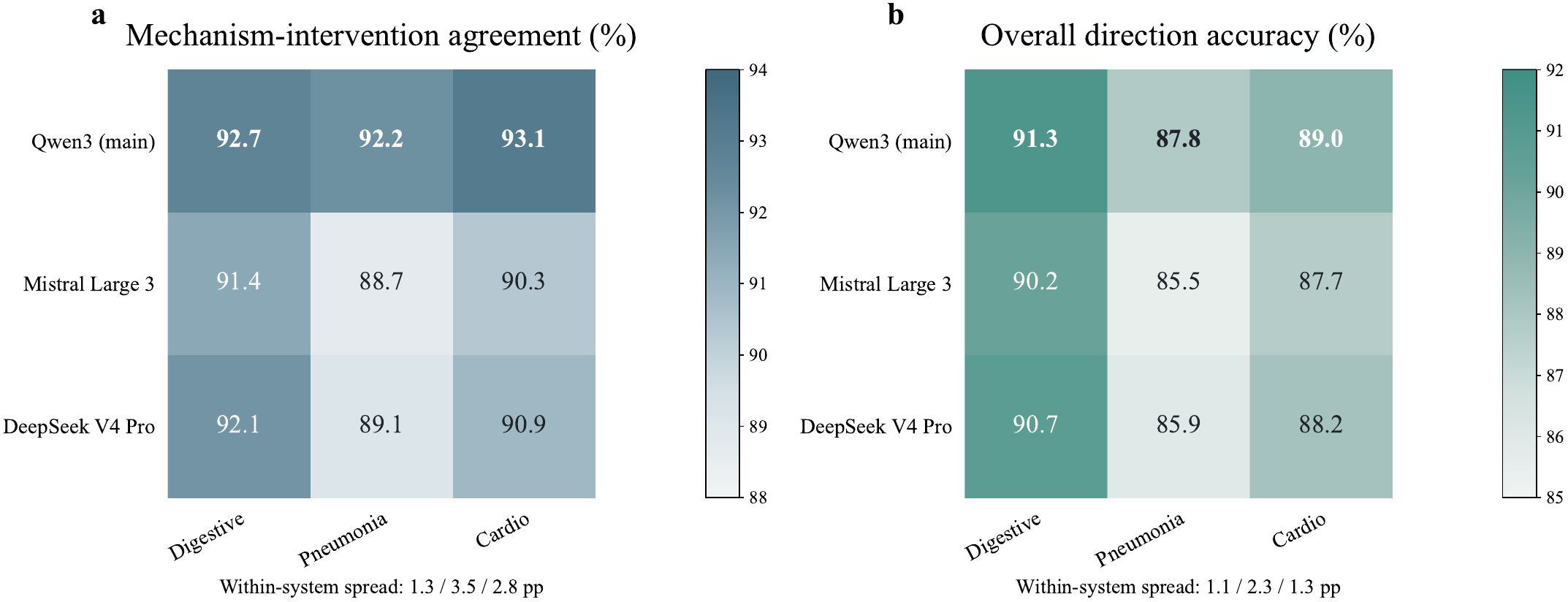}
  \caption{Between-backbone variation on IDAR across the three systems. Qwen3-235B-A22B-Instruct-2507 is the main backbone; Mistral-Large-3-675B-Instruct-2512 and DeepSeek-V4-Pro-0813 are run on the same cases, prompts, and temperature.}
  \label{fig:backbone_robustness}
\end{figure}

\begin{figure}[t]
  \centering
  \includegraphics[width=\linewidth]{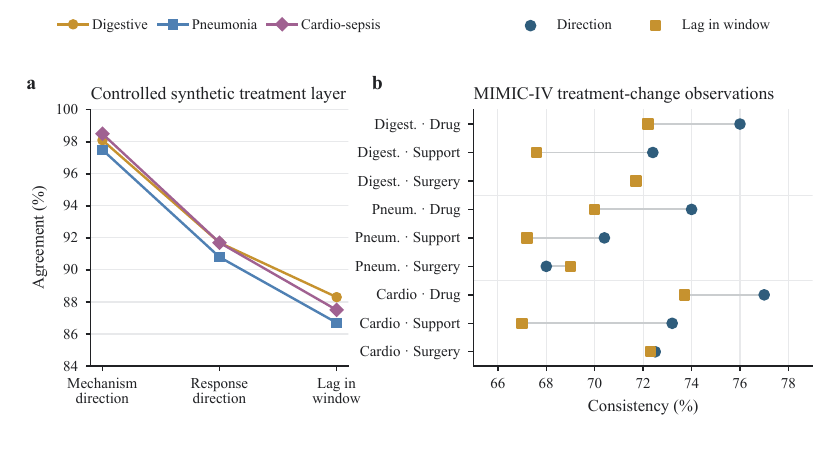}
  \caption{Treatment response and response timing. \textbf{a}, Direction consistency and lag-window agreement on synthetic treatment events. \textbf{b}, Real treatment-change observations stratified by treatment family. Drug changes are most consistent on direction; supportive care is weakest on timing in all three systems.}
  \label{fig:treatment_response}
\end{figure}

\subsubsection{A controlled backbone sweep across all three systems}
\label{sec:backbone_sweep}

All main systems use the same Qwen3 backbone \citep{yang2025qwen3}, while Mistral-Large-3 and DeepSeek-V4-Pro \citep{xu2026deepseek} provide cross-backbone checks. We run every backbone on every mechanism graph with identical cases, prompts, and temperature. The full $3\times3$ sweep in Table~\ref{tab:backbone_sweep} reports the observed between-backbone variation; these values are robustness checks, not a replacement for the topology controls below. The mechanism claim is based on matched graph-bearing interventions, not on comparison with systems for which IDAR is undefined.

\begin{table}[t]
    \centering
    \small
    \setlength\tabcolsep{3pt}
    \renewcommand{\arraystretch}{1.05}
    \begin{tabular}{@{}lccc@{}}
        \toprule
        \textbf{Backbone} & \textbf{Digestive} & \textbf{Pneumonia} & \textbf{Cardio-sepsis} \\
        \midrule
        \multicolumn{4}{l}{\emph{PhysioCircuit edge-IDAR (\%)}} \\
        Qwen3-235B-A22B-Instruct-2507 (main) & \cellcolor{teal!50}\textbf{92.7} & \cellcolor{teal!50}\textbf{92.2} & \cellcolor{teal!50}\textbf{93.1} \\
        Mistral-Large-3-675B-Instruct-2512 & \cellcolor{teal!47}91.4 & \cellcolor{teal!43}88.7 & \cellcolor{teal!45}90.3 \\
        DeepSeek-V4-Pro-0813 & \cellcolor{teal!48}92.1 & \cellcolor{teal!44}89.1 & \cellcolor{teal!46}90.9 \\
        within-system spread (max $-$ min) & 1.3 & 3.5 & 2.8 \\
        \midrule
        \multicolumn{4}{l}{\emph{Global agent input-IDAR (\%)}} \\
        Qwen3-235B-A22B-Instruct-2507 (main) & \cellcolor{gray!18}35.4 & \cellcolor{gray!16}33.7 & \cellcolor{gray!18}36.0 \\
        Mistral-Large-3-675B-Instruct-2512 & \cellcolor{gray!17}34.8 & \cellcolor{gray!16}33.1 & \cellcolor{gray!18}35.5 \\
        DeepSeek-V4-Pro-0813 & \cellcolor{gray!18}35.9 & \cellcolor{gray!17}34.2 & \cellcolor{gray!19}36.6 \\
        within-system spread (max $-$ min) & 1.1 & 1.1 & 1.1 \\
        \midrule
        \multicolumn{4}{l}{\emph{PhysioCircuit prompt-only input-IDAR (\%)}} \\
        Qwen3-235B-A22B-Instruct-2507 (main) & \cellcolor{blue!32}62.1 & \cellcolor{blue!30}59.8 & \cellcolor{blue!34}63.4 \\
        Mistral-Large-3-675B-Instruct-2512 & \cellcolor{blue!31}60.5 & \cellcolor{blue!29}58.1 & \cellcolor{blue!32}61.8 \\
        DeepSeek-V4-Pro-0813 & \cellcolor{blue!31}61.4 & \cellcolor{blue!30}59.0 & \cellcolor{blue!33}62.7 \\
        within-system spread (max $-$ min) & 1.6 & 1.7 & 1.6 \\
        \midrule
        \multicolumn{4}{l}{\emph{Overall direction accuracy (\%)}} \\
        Qwen3-235B-A22B-Instruct-2507 (main) & \cellcolor{green!50}\textbf{91.3} & \cellcolor{green!44}\textbf{87.8} & \cellcolor{green!46}\textbf{89.0} \\
        Mistral-Large-3-675B-Instruct-2512 & \cellcolor{green!48}90.2 & \cellcolor{green!40}85.5 & \cellcolor{green!44}87.7 \\
        DeepSeek-V4-Pro-0813 & \cellcolor{green!49}90.7 & \cellcolor{green!41}85.9 & \cellcolor{green!45}88.2 \\
        within-system spread (max $-$ min) & 1.1 & 2.3 & 1.3 \\
        \bottomrule
    \end{tabular}
    \caption{Cross-backbone robustness sweep. The Qwen row is the main backbone; Mistral and DeepSeek are additional models run on the same cases, prompts, and temperature, with only the backbone changed. The ordering within each channel is stable across backbones: within edge-IDAR, the correct graph remains highest; within input-IDAR, prompt-only exceeds implicit graph learning and the global agent.}
    \label{tab:backbone_sweep}
\end{table}

Figure~\ref{fig:complexity_strata} shows the stratified breakdown. PhysioCircuit gains little on simple cases, where one organ dominates, and most on turning-point cases, where macro transitions require arbitration. Its gain over the global agent rises from 1.2, 0.8, and 0.8 points on simple cases to 13.6, 12.3, and 14.0 points of turning-point F1 on turning-point cases. On conflict cases, PhysioCircuit leads implicit graph learning by 3.1 points on digestive (92.3 vs.\ 89.2) and 3.6 points on cardiovascular (87.6 vs.\ 84.0); its advantage widens further on turning-point and non-monotonic cases, where the answer depends on a mechanism transition instead of state co-occurrence. This is why accuracy stratification and intervention testing are both needed.

\subsubsection{Mechanism intervention and the independence probe}
\label{sec:intervention}

Figure~\ref{fig:mechanism_interventions}(a) shows the matched topology comparison on the digestive system; the headline edge-IDAR and input-IDAR values are reported in Section~\ref{sec:experiments}. These controls all expose a graph-bearing intervention interface, so the comparison asks whether topology and direction determine the response. Table~\ref{tab:intervention_applicability} summarizes both channels and all controls; Table~\ref{tab:idar_decomposition} reports DA, TA, and joint IDAR separately; Table~\ref{tab:intervention_operations} breaks down the six intervention operations; Table~\ref{tab:topology_all_systems} extends the topology controls to all three systems.

Figure~\ref{fig:mechanism_interventions}(b) reports PhysioCircuit's joint IDAR by operation. Delete, replace, reverse, extend-lag, inject-wrong-message, and mask-evidence interventions are evaluated under the same direction-and-window contract. The ordering is stable across systems.

\begin{table}[!ht]
    \centering
    \small
    \setlength\tabcolsep{4pt}
    \renewcommand{\arraystretch}{1.08}
    \begin{tabular}{@{}L{3.4cm} L{2.6cm} L{4.6cm}@{}}
        \toprule
        \textbf{Control} & \textbf{IDAR (\%)} & \textbf{Intervention channel} \\
        \midrule
        \multicolumn{3}{l}{\emph{Edge-IDAR (graph-bearing, digestive system)}} \\
        Correct expert graph & \cellcolor{teal!50}\textbf{92.7} & Named edges; all six operations applicable \\
        Fully connected graph & \cellcolor{teal!36}71.3 & Named edges; extra non-physiological links retained \\
        Random graph & \cellcolor{teal!14}31.2 & Named edges; topology randomized \\
        Reversed graph & \cellcolor{teal!12}27.9 & Named edges; directions reversed \\
        \midrule
        \multicolumn{3}{l}{\emph{Input-IDAR (edge-less, D/P/C)}} \\
        Global single agent & 35.4 / 33.7 / 36.0 & Controlled input-level counterfactual \\
        Organ + optimal voting & 38.2 / 36.5 / 39.1 & Controlled input-level counterfactual \\
        No communication & 37.0 / 35.8 / 38.3 & Controlled input-level counterfactual \\
        Vital Trace-style & 45.2 / 43.8 / 46.1 & Controlled input-level counterfactual \\
        Implicit graph learning & 58.4 / 56.7 / 59.2 & Controlled input-level counterfactual \\
        PhysioCircuit prompt-only & 62.1 / 59.8 / 63.4 & Same LLM and cards; graph propagation disabled \\
        \bottomrule
    \end{tabular}
    \caption{IDAR channels and reported values. Graph-bearing controls use edge-IDAR on named mechanism edits; these rows report the digestive system. Edge-less controls use input-IDAR on the same semantic counterfactual descriptors; their slash-separated values are digestive / pneumonia / cardiovascular-sepsis. The PhysioCircuit prompt-only row disables typed messages and graph propagation while retaining the same LLM and card knowledge, isolating the contribution of the graph channel from model knowledge.}
    \label{tab:intervention_applicability}
\end{table}

\begin{table}[!ht]
    \centering
    \small
    \setlength\tabcolsep{3pt}
    \renewcommand{\arraystretch}{1.06}
    \begin{tabular}{@{}lcccc@{}}
        \toprule
        \textbf{Control} & \textbf{DA (\%)} & \textbf{TA (\%)} & \textbf{Joint IDAR (\%)} & \textbf{95\% CI} \\
        \midrule
        \multicolumn{5}{l}{\emph{Edge-IDAR (digestive, $N=180$ applicable interventions)}} \\
        Correct expert graph & \cellcolor{teal!50}94.1 & \cellcolor{teal!50}95.2 & \cellcolor{teal!50}\textbf{92.7} & [90.1, 94.8] \\
        Fully connected graph & \cellcolor{teal!37}74.2 & \cellcolor{teal!42}82.1 & \cellcolor{teal!36}71.3 & [67.5, 74.8] \\
        Random graph & \cellcolor{teal!20}42.3 & \cellcolor{teal!28}58.7 & \cellcolor{teal!14}31.2 & [27.6, 35.0] \\
        Reversed graph & \cellcolor{teal!15}33.5 & \cellcolor{teal!40}78.4 & \cellcolor{teal!12}27.9 & [24.4, 31.6] \\
        \midrule
        \multicolumn{5}{l}{\emph{Input-IDAR (digestive, $N=180$ descriptors)}} \\
        Global single agent & \cellcolor{blue!22}44.6 & \cellcolor{blue!36}72.3 & \cellcolor{blue!18}35.4 & [31.2, 39.8] \\
        Optimal voting & \cellcolor{blue!24}47.1 & \cellcolor{blue!37}74.0 & \cellcolor{blue!19}38.2 & [33.9, 42.7] \\
        No communication & \cellcolor{blue!23}45.8 & \cellcolor{blue!37}73.5 & \cellcolor{blue!18}37.0 & [32.7, 41.6] \\
        Vital Trace-style & \cellcolor{blue!26}51.8 & \cellcolor{blue!40}79.4 & \cellcolor{blue!23}45.2 & [40.8, 49.7] \\
        Implicit graph learning & \cellcolor{blue!32}63.2 & \cellcolor{blue!43}85.1 & \cellcolor{blue!30}58.4 & [54.1, 62.5] \\
        PhysioCircuit prompt-only & \cellcolor{blue!34}67.3 & \cellcolor{blue!41}82.6 & \cellcolor{blue!31}62.1 & [58.0, 66.0] \\
        \bottomrule
    \end{tabular}
    \caption{Direction Agreement (DA), Timing Agreement (TA), and joint IDAR on the digestive system. DA and TA are reported separately to show whether failures arise from wrong direction or missed timing windows. 95\% bootstrap intervals resample intervention instances. The reversed-graph row keeps TA above the random graph (78.4 vs.\ 58.7) but below the fully connected graph (82.1), while DA falls near the chance level of a three-way choice: the timing channel partially survives and the directional answer loses its information.}
    \label{tab:idar_decomposition}
\end{table}

\begin{table}[!ht]
    \centering
    \small
    \setlength\tabcolsep{3pt}
    \renewcommand{\arraystretch}{1.06}
    \begin{tabular}{@{}lccc@{}}
        \toprule
        \textbf{Intervention operation} & \textbf{Digestive} & \textbf{Pneumonia} & \textbf{Cardio-sepsis} \\
        \midrule
        Delete mechanism & \cellcolor{teal!47}93.2 & \cellcolor{teal!46}92.5 & \cellcolor{teal!48}93.8 \\
        Replace mechanism card & \cellcolor{teal!45}91.8 & \cellcolor{teal!44}91.0 & \cellcolor{teal!46}92.3 \\
        Reverse mechanism direction & \cellcolor{teal!43}90.5 & \cellcolor{teal!42}89.8 & \cellcolor{teal!44}91.2 \\
        Extend lag window & \cellcolor{teal!45}92.1 & \cellcolor{teal!44}91.5 & \cellcolor{teal!46}92.7 \\
        Inject wrong message & \cellcolor{teal!49}94.0 & \cellcolor{teal!47}93.3 & \cellcolor{teal!49}94.4 \\
        Mask source evidence & \cellcolor{teal!49}94.5 & \cellcolor{teal!48}93.9 & \cellcolor{teal!50}95.0 \\
        \midrule
        Average & \cellcolor{teal!46}92.7 & \cellcolor{teal!45}92.0 & \cellcolor{teal!47}93.2 \\
        \bottomrule
    \end{tabular}
    \caption{PhysioCircuit edge-IDAR by intervention operation across the three systems. All six operations use the same direction-and-window scoring contract. Inject-wrong-message and mask-source-evidence are the most faithful because they force the receiving organ to arbitrate. Values are joint DA$\wedge$TA over all applicable interventions ($N=180/210/190$).}
    \label{tab:intervention_operations}
\end{table}

\begin{table}[!ht]
    \centering
    \small
    \setlength\tabcolsep{3pt}
    \renewcommand{\arraystretch}{1.06}
    \begin{tabular}{@{}lccc@{}}
        \toprule
        \textbf{Topology control} & \textbf{Digestive} & \textbf{Pneumonia} & \textbf{Cardio-sepsis} \\
        \midrule
        Correct expert graph & \cellcolor{teal!50}\textbf{92.7} & \cellcolor{teal!50}\textbf{92.2} & \cellcolor{teal!52}\textbf{93.1} \\
        Fully connected graph & \cellcolor{teal!35}71.3 & \cellcolor{teal!34}69.8 & \cellcolor{teal!36}72.5 \\
        Random graph & \cellcolor{teal!12}31.2 & \cellcolor{teal!11}29.8 & \cellcolor{teal!13}32.4 \\
        Reversed graph & \cellcolor{teal!10}27.9 & \cellcolor{teal!9}26.5 & \cellcolor{teal!10}28.7 \\
        \bottomrule
    \end{tabular}
    \caption{Edge-IDAR for graph-bearing topology controls across all three systems. The correct-graph lead over the fully connected graph is stable (20.6--22.4 points), and random and reversed graphs remain near the lower bound in every system.}
    \label{tab:topology_all_systems}
\end{table}

The independence probe injects 18 wrong truth rules. PhysioCircuit retains the medically correct direction on 84.6\%, 81.5\%, and 83.2\% of probe interventions, follows the injected rules on 11.2\%, 13.6\%, and 12.0\%, and produces no directional response on the remainder (4.2\%, 4.9\%, and 4.8\%). Among the interventions that retain the correct direction, 88.1\%, 86.4\%, and 87.5\% also fall within the card-defined lag window, yielding joint probe IDAR of 74.5\%, 70.4\%, and 72.8\%. The truth-retrieval control follows the injected rules on 92.3\% of cases. Source overlap remains below 30\% in all systems. Treatment responses are evaluated separately because treatments enter through organ evidence.

\subsubsection{Ablation analysis}

We ablate mechanism knowledge, graph topology, program gates, and update magnitude on the digestive system; the same ordering holds qualitatively in the other systems. This design separates knowledge, communication, topology, robustness, and long-horizon stability.

Figure~\ref{fig:component_ablations}(a) shows that removing cards costs 4.6 points (91.3 to 86.7), replacing typed messages with free text costs 3.4 points (91.3 to 87.9), and optimal voting over organ states costs 7.8 points (83.5). Adding compute without the graph reaches 84.2 and recovers only part of the gap. These controls separate the contribution of mechanism knowledge, typed communication, macro arbitration, and topology.

Figure~\ref{fig:component_ablations}(b) shows a 10.8-point loss with a random graph (91.3 vs.\ 80.5, paired CI [8.9, 12.7], $p<0.001$), 79.8 accuracy with reversed edges, and 81.6 without communication. A fully connected graph reaches 85.1, recovering part of the gap but remaining 6.2 points below PhysioCircuit. The corresponding edge-IDAR and input-IDAR values are reported in Section~\ref{sec:experiments}.

Figure~\ref{fig:component_ablations}(c) shows the fourteen gates as six atomic checks, edge existence, direction, lag, evidence, duplicate detection, and verdict. Removing the four functional groups: (i) edge existence + duplicate detection, (ii) direction + lag, (iii) evidence, (iv) verdict. This lowers invalid-message interception from 87.6\% to 74.2\%, 68.9\%, 66.1\%, and 70.5\%, respectively. Removing all gates increases error spread from 6.8 to 26.3. The gates therefore constrain error propagation, not only message format.

Figure~\ref{fig:component_ablations}(d) shows that one-step accuracy is nearly flat across the bound, 88.8, 88.7, 88.6, and 88.4 at $\alpha_{\max}=0.3$, 0.6, 0.8, and 1.0, so accuracy does not choose the operating point. The choice rests on long-run behavior: at 0.6, five-step error growth is 0.42 versus 0.89 at 1.0, and direction flips fall from 7.1 to 3.1 per case. A smaller bound of 0.3 under-revises conflict cases, where conflict accuracy drops to 87.5 against 92.3 at 0.6, while a bound of 1.0 accumulates error across steps. We set the operating point at 0.6.

Together, the ablations assign distinct roles to cards, typed messages, topology, gates, and bounded updates: the cards carry knowledge, typed messages support arbitration, topology determines propagation, gates limit error spread, and bounded updates stabilize rollouts. Each removal degrades accuracy, intervention, robustness, or stability. The ablations are centered on the digestive system, so transfer across systems tests whether this decomposition remains useful beyond the source graph.

\subsubsection{Transfer across systems}
\label{sec:transfer}

Figure~\ref{fig:cross_system_transfer}(a) compares zero-shot graph reuse with configuration-only transfer. Rewritten pneumonia and cardiovascular graphs require 127 and 154 configuration lines and reach 87.5\% and 88.7\%, 3.8 and 2.6 points below the digestive source. Zero-shot losses range from 3.9 to 11.2 points and track organ overlap: the digestive graph shares no organ with either six-organ system, while the pneumonia and cardiovascular graphs share heart and kidney structure. Panel B shows larger reuse losses for implicit graph learning (7.8--17.6 points vs.\ 3.9--11.2 for PhysioCircuit); PhysioCircuit leads in all six directions by 1.8--7.2 points. The result supports mechanism cards as reusable relations, while also motivating validation on records outside the generated benchmark.

\subsubsection{The real longitudinal pillar}

\begin{figure}[!ht]
  \centering
  \includegraphics[width=\linewidth]{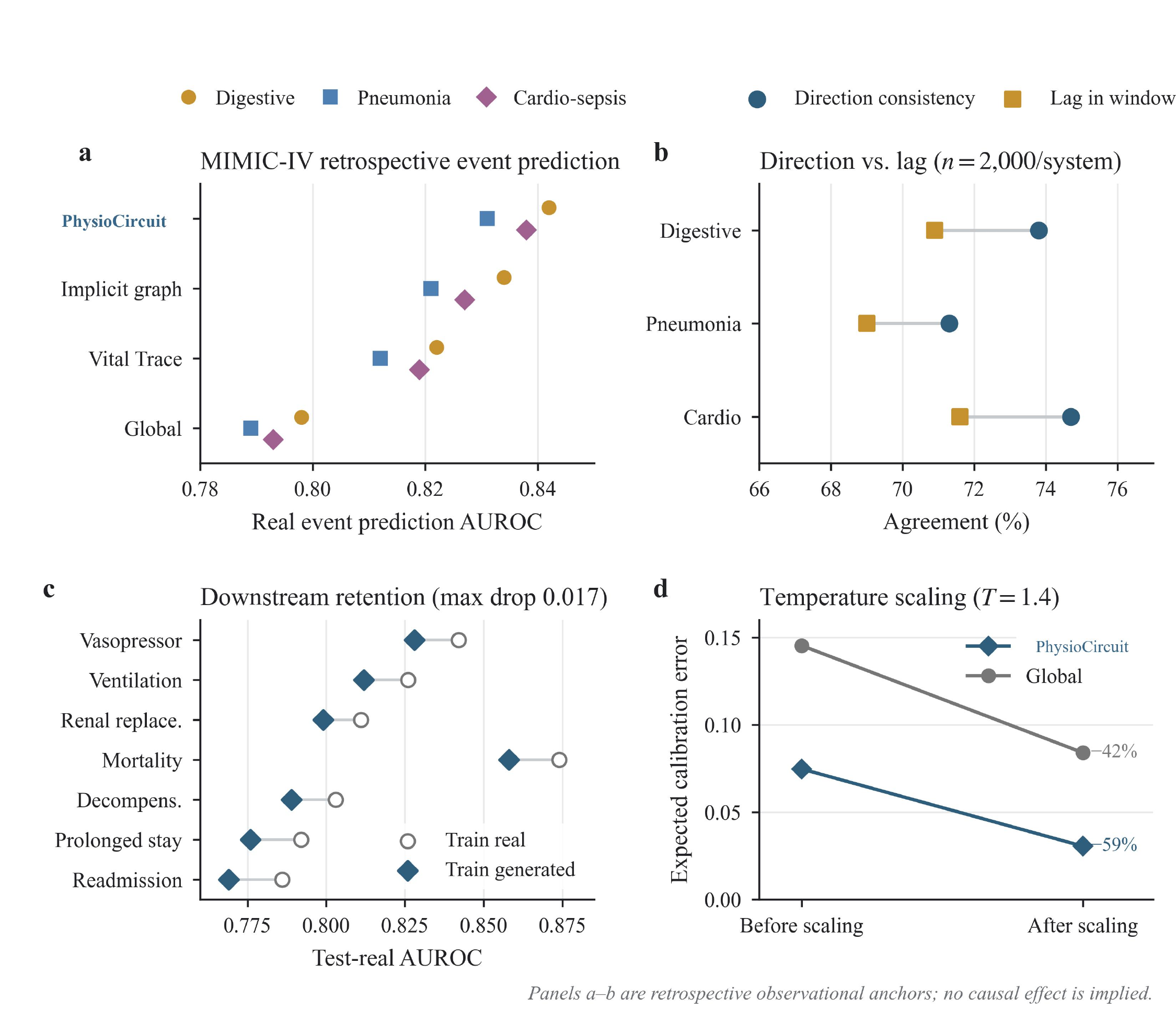}
  \caption{Real-data and downstream validation. \textbf{a}, Retrospective MIMIC-IV event-prediction AUROC. \textbf{b}, Direction consistency and lag-window agreement on 2,000 treatment-change observations per system. \textbf{c}, AUROC of models trained on real versus generated trajectories and evaluated on the same real test set. \textbf{d}, Overall expected calibration error before and after temperature scaling.}
  \label{fig:real_world_validation}
\end{figure}

On 600 ICU admissions, PhysioCircuit reaches AUROC 0.842, 0.831, and 0.838 for real-event prediction, higher than implicit graph learning (0.834, 0.821, 0.827), Vital Trace-style reasoning (0.822, 0.812, 0.819), and the global agent (0.798, 0.789, 0.793) (Figure~\ref{fig:real_world_validation}(a,b)). Mechanism-direction consistency is 73.8\%, 71.3\%, and 74.7\%, with timing at 70.9\%, 69.0\%, and 71.6\%. These real scores sit below synthetic IDAR because co-interventions go unrecorded and timestamps are coarser, yet the ordering holds.

\subsubsection{Downstream task retention}

Figure~\ref{fig:real_world_validation}(c) evaluates seven ICU tasks with admission-disjoint train, validation, and test splits (300/100/200). The validation split is used only to select the temperature-scaling parameter $T$; no admission appears in more than one split. Models trained on generated or real trajectories are evaluated on the same real test set, holding the target distribution fixed. The average AUROC drop is 0.015 and the largest is 0.017 on unplanned readmission; ICU mortality drops 0.016 from an upper bound of 0.874. The generated trajectories retain task-relevant information, and the expert evaluation supplies the clinical check.

\subsubsection{Uncertainty calibration}
\label{sec:calibration}

Figure~\ref{fig:real_world_validation}(d) reports ECE over 3,984 macro predictions using post-hoc temperature scaling \citep{Guo2017calibration}. Temperature scaling with $T=1.4$ reduces PhysioCircuit ECE from 0.0747 to 0.0305, a 59\% reduction; the global agent changes from 0.1453 to 0.0840. Baseline and recovery phases start near calibration (0.0049 and 0.0057), while conflict-emerges and decompensation reach 0.0141 and 0.0153. Residual overconfidence therefore concentrates where organ evidence conflicts and the framework must arbitrate. Calibration shows both the benefit of post-hoc correction and the phase-specific risk that remains.

\subsubsection{Cost and the zero-training comparison}
\label{sec:cost}

Table~\ref{tab:cost} compares the training burden and extension cost. The Organ-Agents figures are quoted from its report: nine systems, two training stages, one fixed backbone, and retraining when the system changes. PhysioCircuit uses no training, adds systems through mechanism cards and configuration, and costs 41.7K--42.4K tokens per case in the measured settings. Adding pneumonia and cardiovascular-sepsis requires 14 and 12 cards plus 127 and 154 configuration lines. Gates intercept 87.6\% of invalid messages, retain 82.4\% of valid messages, and reduce effective-correction cost by 41.3\% relative to free-text coordination, which spends 71.0K tokens per case on the same task.

\begin{table}[t]
    \centering
    \small
    \setlength\tabcolsep{3pt}
    \renewcommand{\arraystretch}{1.06}
    \begin{tabular}{@{}L{3.7cm} L{4.0cm} L{5.3cm}@{}}
        \toprule
        \textbf{Item} & \textbf{Organ-Agents} & \textbf{PhysioCircuit} \\
        \midrule
        Training stages & 2 (supervised fine-tuning, then reinforcement-guided coordination) & 0 \\
        Physiological systems covered & 9 & 3 \\
        Backbone models & one fixed backbone & one main + two cross-backbone checks \\
        Retraining on a new system & required & not required \\
        Cost of a new system & full training run & 127 and 154 configuration lines, plus 14 and 12 mechanism cards \\
        Synthetic cases per system & not applicable, trained on real cohorts & 120 \\
        Tokens per case & not reported & 41.7K for three organs, 42.4K for six organs \\
        \bottomrule
    \end{tabular}
    \caption{Cost comparison. The Organ-Agents column quotes the figures reported in its own paper; the PhysioCircuit column reports measured values from this work.}
    \label{tab:cost}
\end{table}

\end{document}